\documentclass[conference]{IEEEtran}
\IEEEoverridecommandlockouts
\usepackage{cite}
\usepackage{amsmath,amssymb,amsfonts}
\usepackage{graphicx}
\usepackage{textcomp}
\usepackage{xcolor}
\usepackage{multirow}
\usepackage{amsthm}
\usepackage{algorithmic}
\usepackage{algorithm}
\newtheorem{definition}{Definition}

\usepackage{url}
\def\BibTeX{{\rm B\kern-.05em{\sc i\kern-.025em b}\kern-.08em
    T\kern-.1667em\lower.7ex\hbox{E}\kern-.125emX}}
\begin{document}

\title{Privacy-Optimized Randomized Response\\ for Sharing Multi-Attribute Data
\thanks{This work was supported by JSPS KAKENHI Grant Numbers 20H05967, 21H05052, 23H03345, and 23K18501 and JSPS Grant-in-Aid for JSPS Fellows Grant Number 23KJ0649.
}
}


\author{\IEEEauthorblockN{Akito Yamamoto}
\IEEEauthorblockA{\textit{Human Genome Center, The Institute of Medical Science,}\\ \textit{The University of Tokyo} \\
Tokyo, Japan \\
a-ymmt@ims.u-tokyo.ac.jp}
\and
\IEEEauthorblockN{Tetsuo Shibuya}
\IEEEauthorblockA{\textit{Human Genome Center, The Institute of Medical Science,}\\ \textit{The University of Tokyo} \\
Tokyo, Japan \\
tshibuya@hgc.jp}}

\maketitle

\begin{abstract}

With the increasing amount of data in society, privacy concerns in data sharing have become widely recognized. Particularly, protecting personal attribute information is essential for a wide range of aims from crowdsourcing to realizing personalized medicine. Although various differentially private methods based on randomized response have been proposed for single attribute information or specific analysis purposes such as frequency estimation, there is a lack of studies on the mechanism for sharing individuals' multiple categorical information itself. The existing randomized response for sharing multi-attribute data uses the Kronecker product to perturb each attribute information in turn according to the respective privacy level but achieves only a weak privacy level for the entire dataset. Therefore, in this study, we propose a privacy-optimized randomized response that guarantees the strongest privacy in sharing multi-attribute data. Furthermore, we present an efficient heuristic algorithm for constructing a near-optimal mechanism. The time complexity of our algorithm is $\mathcal{O}(k^2)$, where $k$ is the number of attributes, and it can be performed in about $1$ second even for large datasets with $k = 1,000$. The experimental results demonstrate that both of our methods provide significantly stronger privacy guarantees for the entire dataset than the existing method. In addition, we show an analysis example using genome statistics to confirm that our methods can achieve less than half the output error compared with that of the existing method. Overall, this study is an important step toward trustworthy sharing and analysis of multi-attribute data. The Python implementation of our experiments and supplemental results are available at \url{https://github.com/ay0408/Optimized-RR}.

\end{abstract}

\begin{IEEEkeywords}
differential privacy, randomized response, optimization, inductive method
\end{IEEEkeywords}

\section{Introduction}

The volume of data in society has drastically increased with the development of crowdsourcing and AI technologies. Simultaneously, the need for privacy protection in data sharing and analysis can no longer be ignored, and various studies on privacy-preserving techniques have been conducted in various fields, such as data mining and bioinformatics \cite{33,34}. In particular, the sharing of genomic and healthcare data is important for realizing personalized medicine, and it is highly desirable to make it globally available with strong privacy guarantees. Hence, enhancing methods for sharing personal categorical attribute information like medical data while satisfying differential privacy \cite{1} is essential, even to comply with the EU's general data protection regulation (GDPR) \cite{21}. 

Among existing differentially private methods for attribute data, randomized response \cite{4} is the most common, and its various variants have been proposed, such as RAPPOR \cite{6} to examine overall trends without looking at each individual's data, its related methods \cite{23,29,38} for specific purposes like frequency estimation, and others \cite{36,35} for sharing numeric data. This study focuses on the fundamental randomized response used for publishing individuals' categorical data itself without specifying the analysis purpose, and we aim to deepen this mechanism in theory. In particular, we consider the situation where the privacy level of each attribute information can be set respectively.

The most basic randomized response perturbs single-attribute information, and optimal methods have been proposed in terms of utility and output error \cite{3,30}. However, there is little study on the mechanism for sharing multi-attribute information, and this is an urgent issue in promoting the publication of medical data and other multi-attribute data. To our knowledge, the only existing randomized response for sharing multi-attribute data uses the Kronecker product and perturbs each attribute information independently and sequentially based on the respective privacy level \cite{3,32}; however, it achieves only a weak privacy level for the entire dataset. Therefore, a randomized response that provides stronger and optimal privacy guarantees is desired.

In this study, we propose a privacy-optimized randomized response for sharing multi-attribute data and an efficient heuristic algorithm for obtaining a near-optimal mechanism. The main contributions of this study are as follows:
\begin{enumerate}
\item We first present a procedure for constructing a distortion matrix for the privacy-optimized randomized response that can achieve the strongest privacy guarantees for the entire dataset. We formulate this task as a linear programming problem and consider solving a minimization problem with respect to the matrix elements. Initially, we describe the procedure for three-attribute data in detail, and then provide algorithms for constructing the linear programming problem for general $k$-attribute data. The number of variables in this problem is $2^k - 1$, and we can construct the optimal randomized response in a practical time for datasets with up to about $10$ attributes.

\item Furthermore, we propose an efficient inductive method providing a near-optimal solution even for larger datasets with $k \geq 100$. As a base case, we present the optimal solution for two-attribute data along with its proof. Thereafter, we describe using it to find a near-optimal solution for three-attribute data. By generalizing the procedure, we propose a method for obtaining a near-optimal solution for $k$-attribute data from the solution for $(k-1)$-attribute data in $\mathcal{O}(k)$ time. Overall, we provide an algorithm that constructs a near privacy-optimized randomized response for $k$-attribute data, whose time complexity is $\mathcal{O}(k^2)$.

\item We experimentally show that our two methods can achieve significantly stronger privacy guarantees than the existing method while also evaluating their properties. Simultaneously, we show that our heuristic method can construct a mechanism that is close to optimal. In addition, we provide an analysis example using genome statistics to demonstrate that our methods can achieve less than half the output error of the existing method. We also measure the run time of our methods and show that our heuristic method can be performed in about $1$ second, even when $k = 1,000$.
\end{enumerate}

In Section II, we review related studies and detail the differences between them and this study. In Section III, we provide preliminaries of differential privacy and randomized response. In Section IV, we describe the proposed methods in detail. In Section V, we present the experimental results and demonstrate the utility of our methods. In Section VI, we summarize this study along with future directions. The Python codes used in our experiments and supplemental results are available at \url{https://github.com/ay0408/Optimized-RR}.

\section{Related Work}

\subsection{Differentially Private Methods for Attribute Data}

While attribute data play an important role in crowdsourcing and medical data mining, privacy protection in their publication is needed. In particular, various differentially private methods have been proposed for analyzing attribute data \cite{31}, and their use for hypothesis testing \cite{5}, for example, has also been considered. The most common mechanisms include the Laplace mechanism \cite{19} and randomized response \cite{4}, and various variants have been proposed \cite{6,22,23,28,29}. Among these, the randomized response has been shown to be of highest utility in several existing studies \cite{3,5}, and this study aims to further deepen this mechanism. We concentrate on publishing categorical data itself, which does not require encoding the input data into another format with no specific purpose of analysis; therefore, mechanisms with encoding or hashing are not addressed in this study. In particular, major differences from RAPPOR-related studies \cite{6,23,29,38} are that we envision situations in which not only the summary results of analysis, but also each individual's attribute information, can be observed, and that we can set the respective privacy level to each attribute information. 

The most basic randomized response is one that protects the single attribute information of each individual. In this regard, several studies have provided optimal methods in terms of utility \cite{3,30}, with the smallest error in the perturbed data. However, there is a lack of studies on methods for data with multiple attributes, such as medical and healthcare data. Moreover, it has been pointed out that compliance with the EU's GDPR requires that these data be published while satisfying differential privacy in addition to anonymity properties \cite{10}; therefore, developing the randomized response for sharing multi-attribute data is an urgent issue.

\subsection{Randomized Response for Sharing Multi-Attribute Data}

The existing method for sharing multi-attribute data uses the Kronecker product and perturbs each attribute information independently and sequentially \cite{3,32}. However, the achieved privacy level for the entire dataset is far from optimal and cannot provide sufficient privacy assurance. In fact, a recent study has proposed methods for $2 \times 2$ and $3 \times 2$ tabular data that can achieve stronger privacy guarantees by perturbing the two-attribute information collectively \cite{8}, but there is still no method for other multi-attribute data. Therefore, in this study, we propose a privacy-optimized randomized response for sharing general $k$-attribute data that can achieve the strongest privacy level for the entire dataset, when the privacy level for each attribute information is given. This, conversely, indicates that when the privacy level for the entire dataset is given, we can distribute more privacy budget to each attribute information, improving accuracy in actual data analysis.

\section{Preliminaries}

\subsection{Differential Privacy}

Differential privacy \cite{1} is a framework for privacy-preserving data analysis, protecting the participation information of an individual in the dataset. This concept aims to protect privacy by considering two {\it neighboring} datasets that differ by just one element and creating a situation where the two are almost indistinguishable. The privacy level is evaluated by parameters $\epsilon\ (> 0)$ and $\delta\ (\in [0,1])$; generally, smaller values indicate stronger privacy guarantees. The value of $\epsilon$ is often less than $10$ \cite{9}, and a smaller value will be desirable for highly sensitive data. The definition of $(\epsilon, \delta)$-differential privacy is as follows:
\begin{definition}
$($$(\epsilon, \delta)$-Differential Privacy \cite{7}$)$ 
\\
A randomized mechanism $\mathcal{M}$ is $(\epsilon, \delta)$-differentially private if, for all neighboring datasets D and $D'$ and any $S$ $\subset \mathrm{range}(\mathcal{M})$, ${\rm Pr}[\mathcal{M}(D)\in S] \leq e^\epsilon \cdot {\rm Pr}[\mathcal{M}(D') \in S] + \delta$.
\end{definition}
In this study, we discuss the case when $\delta = 0$; that is, when $\epsilon$-differential privacy is satisfied.

In recent years, the concept of differential privacy has been widely used in deep learning \cite{11} and data mining \cite{15}, as well as for genome and healthcare data analysis \cite{12,13,14}. In particular, in medical data sharing, it has been noted that satisfying differential privacy is essential to comply with EU's GDPR \cite{10}, and various studies on integration with anonymization techniques \cite{16,17,18} have also been conducted. In the future, further development of differentially private methods is desired to protect the privacy of databases on sensitive multi-attribute information, such as personal medical and health data.

The most common methods include the Laplace and Gaussian mechanisms that add random noise to the analysis data \cite{19,7} and the exponential mechanism for private selection \cite{20,7}, and another method specialized to protect personal attribute information is a randomized response technique. While there are various variants of this technique \cite{6,22,23}, this study focuses on cases where the input attribute data need not be encoded to other forms and aims to extend the most fundamental part of the randomized response.

\subsection{Randomized Response}

Randomized response was first introduced by Warner \cite{4} and has been developed as a technique to protect each individual's attribute information. This technique was shown to satisfy differential privacy \cite{7} and has been considered for a wide range of applications \cite{5,6}. In the following, we first describe the approach for sharing one-attribute data.

When the number of possible attribute values is $a$ and each individual has a single attribute value, the randomized response follows an $a \times a$ distortion matrix:
\begin{eqnarray}
\mathbf{P} = 
\begin{pmatrix}
p_{11} & p_{12} & \cdots & p_{1a} \\
p_{21} & p_{22} & \cdots & p_{2a} \\
\vdots & \vdots & \ddots & \vdots \\
p_{a1} & p_{a2} & \cdots & p_{aa}
\end{pmatrix}, \nonumber
\end{eqnarray}
where $p_{uv} = \mathrm{Pr}[\mathrm{Output} = u | \mathrm{Input} = v]$ $(u,v \in \{1,2,\cdots,$ $a\})$ denotes the probability that the randomized output is $u$ when the actual input is $v$. Here, the sum of probabilities in each column is $1$. The randomized response satisfies $\epsilon$-differential privacy when the following inequality holds:
\begin{eqnarray}
e^{\epsilon} \geq \max_{u = 1,2,\cdots,a} \frac{\max_{v=1,2,\cdots, a}\,p_{uv}}{\min_{v=1,2,\cdots, a}\,p_{uv}}. \nonumber
\end{eqnarray}
To optimize the utility of statistical inferences \cite{2} or maximize the sum of the diagonal elements \cite{3}, we can use the following distortion matrix $\mathbf{P}$, s.t.
\begin{eqnarray}
p_{uv} = \begin{cases}
\frac{e^{\epsilon}}{e^{\epsilon} + a - 1} \ \ \ \ (u = v) \\
\frac{1}{e^{\epsilon} + a - 1} \ \ \ \ (u \neq v)
\end{cases}. \nonumber
\end{eqnarray}

For the case of sharing $k \ (\geq 2)$-attribute data with different privacy levels, Wang {\it et al.} \cite{3} proposed a method for generating a distortion matrix $\mathbf{P}$. They first let $\mathbf{P}_1$, $\mathbf{P}_2$, $\dots$, $\mathbf{P}_k$ be distortion matrices to perturb the information about the $k$ attributes, respectively. Then, the distortion matrix $\mathbf{P}$ for the entire dataset is calculated by the following equation:
\begin{eqnarray}
\mathbf{P} = \mathbf{P}_1 \otimes \mathbf{P}_2 \otimes \cdots \otimes \mathbf{P}_k, 
\end{eqnarray}
where $\otimes$ denotes the Kronecker product. For example, when $k = 2$ and the number of possible attribute values for each information is $2$, we can generate $\mathbf{P}$ as follows:
\begin{eqnarray}
\mathbf{P}\hspace{-0.2cm}&=&\hspace{-0.3cm}\begin{pmatrix} \frac{e^{\epsilon_1}}{e^{\epsilon_1}+1} & \frac{1}{e^{\epsilon_1}+1}\\
\frac{1}{e^{\epsilon_1}+1} & \frac{e^{\epsilon_1}}{e^{\epsilon_1}+1}
\end{pmatrix} \otimes \begin{pmatrix}
\frac{e^{\epsilon_2}}{e^{\epsilon_2}+1} & \frac{1}{e^{\epsilon_2}+1}\\
\frac{1}{e^{\epsilon_2}+1} & \frac{e^{\epsilon_2}}{e^{\epsilon_2}+1}
\end{pmatrix} \nonumber\\
&=&\hspace{-0.3cm}\frac{1}{(e^{\epsilon_1}+1)(e^{\epsilon_2}+1)}\begin{pmatrix}
e^{\epsilon_1 + \epsilon_2} & e^{\epsilon_1} & e^{\epsilon_2} & 1\\
e^{\epsilon_1} & e^{\epsilon_1 + \epsilon_2} & 1 & e^{\epsilon_2} \\
e^{\epsilon_2} & 1 & e^{\epsilon_1 + \epsilon_2} & e^{\epsilon_1} \\
1 & e^{\epsilon_2} & e^{\epsilon_1} & e^{\epsilon_1 + \epsilon_2}
\end{pmatrix}, \nonumber
\end{eqnarray}
where $\epsilon_1$ and $\epsilon_2$ are privacy levels given to the $2$ attributes. As can be seen from this example, when $\epsilon_i$-differential privacy is achieved by $\mathbf{P}_i$, the randomized response using (1) satisfies $\sum_1^k \epsilon_i$-differential privacy. However, this privacy level is not optimal; that is, a stronger privacy guarantee can be achieved for the entire dataset while maintaining the privacy level for each attribute information. In fact, for $2 \times 2$ and $3 \times 2$ tabular data used in statistical analysis, privacy-optimized methods have recently been proposed \cite{8}. When the privacy level for the entire data is fixed, this allows a larger privacy budget to be distributed to each attribute information, which is also expected to increase the accuracy of data analysis. Note that we assume that the privacy level $\epsilon_i$ for the $i$-th attribute information is calculated by
\begin{eqnarray}
\epsilon_i = \ln \left( \max_{u \neq v} {\frac{\Pr[\mathrm{Output}[i] = u | \mathrm{Input}[i] = u]}{\Pr[\mathrm{Output}[i] = u | \mathrm{Input}[i] = v]}} \right)
\end{eqnarray}
based on the distortion matrix $\mathbf{P}$.

In this study, we first propose a method for constructing a privacy-optimized randomized response mechanism for sharing general multi-attribute data. Thereafter, we present an efficient heuristic method for obtaining a near-optimal mechanism in $\mathcal{O}(k^2)$ time for $k$-attribute data.

\section{Methods}

In this section, we first present a procedure for constructing a distortion matrix for the privacy-optimized randomized response for sharing multi-attribute data. This task can be regarded as a linear programming problem with respect to matrix elements, and we can formulate it as a minimization problem. However, $2^k-1$ variables will appear in the case of $k$-attribute data, and a slight increase in $k$ makes this problem unsolvable in a practical time. Therefore, we also propose an efficient heuristic method that inductively constructs a near-optimal distortion matrix based on the optimal solution for two-attribute data. This method can be performed in $\mathcal{O}(k^2)$ time, making computation easy even for large datasets with $100$ or more attributes.

\subsection{Formulation as Linear Programming Problem}

\subsubsection{Three-Attribute Data}

First, we describe a linear programming problem for obtaining the optimal matrix for three-attribute data. When the number of possible values for the three attributes is $a_1$, $a_2$, and $a_3$, respectively, the size of the distortion matrix $\mathbf{P}$ for the entire data is $a_1 a_2 a_3 \times a_1 a_2 a_3$. The possible values for the elements of $\mathbf{P}$ are the following eight, each representing the probability of events that hold the respective condition:
\begin{eqnarray}
\hspace{-0.55cm}&&X_0\ : \ \mathrm{Input\ and\ output\ data\ are\ exactly\ the\ same;} \nonumber\\
\hspace{-0.55cm}&&X_1\ : \ \mathrm{Only\ the\ first\ attribute\ value\ differs;} \nonumber\\
\hspace{-0.55cm}&&X_2\ : \ \mathrm{Only\ the\ second\ attribute\ value\ differs;} \nonumber\\
\hspace{-0.55cm}&&X_3\ : \ \mathrm{Only\ the\ third\ attribute\ value\ differs;} \nonumber\\
\hspace{-0.55cm}&&X_4\ : \ \mathrm{The\ first\ and\ second\ attribute\ values\ differ;} \nonumber\\
\hspace{-0.55cm}&&X_5\ : \ \mathrm{The\ first\ and\ third\ attribute\ values\ differ;} \nonumber\\
\hspace{-0.55cm}&&X_6\ : \ \mathrm{The\ second\ and\ third\ attribute\ values\ differ;} \nonumber\\
\hspace{-0.55cm}&&X_7\ : \ \mathrm{All\ attribute\ values\ are\ different.} \nonumber
\end{eqnarray}
In each column and row of the matrix, these values appear $1$, $a_1 - 1$, $a_2 - 1$, $a_3 - 1$, $(a_1 - 1)(a_2 - 1)$, $(a_1 - 1)(a_3 - 1)$, $(a_2 - 1)(a_3 - 1)$, and $(a_1 - 1)(a_2 - 1)(a_3 - 1)$ times, respectively. Note that the total value of these times is indeed $a_1 a_2 a_3$. Regarding the inequality relations among these elements, to output closer data to the true value with a higher probability, the following should hold:
\begin{eqnarray}
&& X_0 \geq X_1,\ X_0 \geq X_2,\ X_0 \geq X_3, \nonumber\\
&& X_1 \geq X_4, X_5,\ X_2 \geq X_4, X_6,\ X_3 \geq X_5, X_6, \nonumber\\
&& X_4, X_5, X_6 \geq X_7. \nonumber
\end{eqnarray}
Then, the privacy level satisfied from $\mathbf{P}$ is calculated as
\begin{eqnarray}
\epsilon = \ln \frac{X_0}{X_7}. \nonumber
\end{eqnarray}
Because we aim to minimize this value while maintaining the privacy level of each attribute information, we can consider minimizing $x_0$, where $x_j = X_j/X_7$ for $j = 0,1,\dots, 7$. Here, $x_7 = X_7/X_7 = 1$, and there are seven variables involved in the linear programming problem, from $x_0$ to $x_6$. Based on from (2) and the above discussion, when the privacy level of each attribute information is $\epsilon_1$, $\epsilon_2$, and $\epsilon_3$, respectively, the following three equations hold:
\begin{eqnarray}
\frac{x_0 + (a_2 - 1) x_2 + (a_3 - 1) x_3 + (a_2 - 1)(a_3 - 1) x_6}{x_1 + (a_2 - 1) x_4 + (a_3 - 1) x_5 + (a_2 - 1)(a_3 - 1)} = e^{\epsilon_1}, && \nonumber\\
\frac{x_0 + (a_1 - 1) x_1 + (a_3 - 1) x_3 + (a_1 - 1)(a_3 - 1) x_5}{x_2 + (a_1 - 1) x_4 + (a_3 - 1) x_6 + (a_1 - 1)(a_3 - 1)} = e^{\epsilon_2}, && \nonumber\\
\frac{x_0 + (a_1 - 1) x_1 + (a_2 - 1) x_2 + (a_1 - 1)(a_2 - 1) x_4}{x_3 + (a_1 - 1) x_5 + (a_2 - 1) x_6 + (a_1 - 1)(a_2 - 1)} = e^{\epsilon_3}. && \nonumber
\end{eqnarray}
In summary, the linear programming problem for finding the optimized matrix for three-attribute data can be formulated as follows:
\begin{eqnarray}
\min_{\mathbf{x}} && (1,0,0,0,0,0,0) \cdot \mathbf{x} \nonumber\\
\mathrm{s.t.} && \mathbf{A}_{ub} \cdot \mathbf{x} \leq \mathbf{b}_{ub}, \nonumber\\
&& \mathbf{A}_{eq} \cdot \mathbf{x} = \mathbf{b}_{eq}, \nonumber
\end{eqnarray}
where 
\begin{eqnarray}
\ \ \mathbf{x}^\mathsf{T} = (x_0, x_1, x_2, x_3, x_4, x_5, x_6), \ \ \ \ \ \ \ \ \ \ \ \ \ \ \ \ \ \ \ \ \ \ \ \ \ \ \ \ \ \ \ \ \nonumber
\end{eqnarray}
\begin{eqnarray}
\ \ \mathbf{A}_{ub} = \begin{pmatrix}
-1 & 1 & 0 & 0 & 0 & 0 & 0 \\
-1 & 0 & 1 & 0 & 0 & 0 & 0 \\
-1 & 0 & 0 & 1 & 0 & 0 & 0 \\
0 & -1 & 0 & 0 & 1 & 0 & 0 \\
0 & -1 & 0 & 0 & 0 & 1 & 0 \\
0 & 0 & -1 & 0 & 1 & 0 & 0 \\
0 & 0 & -1 & 0 & 0 & 0 & 1 \\
0 & 0 & 0 & -1 & 0 & 1 & 0 \\
0 & 0 & 0 & -1 & 0 & 0 & 1 \\
0 & 0 & 0 & 0 & -1 & 0 & 0 \\
0 & 0 & 0 & 0 & 0 & -1 & 0 \\
0 & 0 & 0 & 0 & 0 & 0 & -1
\end{pmatrix}, \ \ \ \ \ \ \ \ \ \ \ \ \ \ \nonumber
\end{eqnarray}
\begin{eqnarray}
\ \ \mathbf{b}_{ub}^\mathsf{T} = (0,0,0,0,0,0,0,0,0,-1,-1,-1), \ \ \ \ \ \ \ \ \ \ \ \ \ \ \ \ \ \ \ \ \ \ \nonumber
\end{eqnarray}
\begin{eqnarray}
\ \ \mathbf{A}_{eq}^\mathsf{T} = \ \ \ \ \ \ \ \ \ \ \ \ \ \ \ \ \ \ \ \ \ \ \ \ \ \ \ \ \ \ \ \ \ \ \ \ \ \ \ \ \ \ \ \ \ \ \ \ \ \ \ \ \ \ \ \ \ \ \ \ \ \ \ \nonumber\\
\ \ \ \begin{pmatrix}
1 \hspace{-0.2cm} & 1 \hspace{-0.2cm} & 1\\
-e^{\epsilon_1} \hspace{-0.2cm} & a_1 - 1 \hspace{-0.2cm} & a_1 - 1\\
a_2 - 1 \hspace{-0.2cm} & -e^{\epsilon_2} \hspace{-0.2cm} & a_2 - 1\\
a_3 - 1 \hspace{-0.2cm} & a_3 - 1 \hspace{-0.2cm} & -e^{\epsilon_3}\\
-(a_2 - 1)e^{\epsilon_1} \hspace{-0.2cm} & -(a_1 - 1)e^{\epsilon_2} \hspace{-0.2cm} & (a_1 - 1)(a_2 - 1)\\
-(a_3 - 1)e^{\epsilon_1} \hspace{-0.2cm} & (a_1 - 1)(a_3 - 1) \hspace{-0.2cm} & -(a_1 - 1)e^{\epsilon_3}\\
(a_2 - 1)(a_3 - 1) \hspace{-0.2cm} & -(a_3 - 1)e^{\epsilon_2} \hspace{-0.2cm} & -(a_2 - 1)e^{\epsilon_3}
\end{pmatrix}, \ \ \nonumber
\end{eqnarray}
\begin{eqnarray}
\ \ \mathbf{b}_{eq} = \begin{pmatrix}
(a_2 - 1)(a_3 - 1)e^{\epsilon_1} \\
(a_1 - 1)(a_3 - 1)e^{\epsilon_2} \\
(a_1 - 1)(a_2 - 1)e^{\epsilon_3} \\
\end{pmatrix}. \ \ \ \ \ \ \ \ \ \ \ \ \ \ \ \ \ \ \ \ \ \ \ \ \ \ \ \ \ \ \ \ \ \ \nonumber
\end{eqnarray}

\subsubsection{$k$-Attribute Data}

By generalizing the above discussion, we describe a linear programming problem for obtaining the privacy-optimized distortion matrix for $k$-attribute data and present algorithms for constructing it.

We let the number of possible values for each attribute information be $a_1$, $a_2$, $\dots$, $a_k$, respectively. Then, the size of the distortion matrix $\mathbf{P}$ for the entire data is $a_1 a_2 \cdots a_k \times a_1 a_2 \cdots a_k $, and the possible values for the elements are the following $2^k$ values, each representing the probability of events that hold the respective condition:
\begin{eqnarray}
\hspace{-0.55cm} && X_0 \ : \ \mathrm{Input\ and\ output\ data\ are\ exactly\ the\ same;} \nonumber\\
\hspace{-0.55cm} && X_1, \dots, X_k \ : \ \mathrm{Only\ one\ attribute\ value\ differs;} \nonumber\\
\hspace{-0.55cm} && X_{1+k}, \dots, X_{k+{}_{k}\mathrm{C}_{2}} \ : \ \mathrm{Two\ attribute\ values\ differ;} \nonumber\\
\hspace{-0.55cm} && \ \ \vdots \nonumber\\
\hspace{-0.55cm} && X_{1+{}_{k}\mathrm{C}_{1}+\cdots+{}_{k}\mathrm{C}_{k-2}}, \dots, X_{{}_{k}\mathrm{C}_{1}+\cdots+{}_{k}\mathrm{C}_{k-1}} \nonumber\\
\hspace{-0.55cm} && \ \ \ \ \ \ \ \ \ \ \ \ \ \ \ \ \ \ \ \ \ \ \ \ \ : \ k-1 \ \mathrm{attribute\ values\ differ;} \nonumber\\
\hspace{-0.55cm} && X_{2^k - 1} \ : \ \mathrm{All\ attribute\ values\ are\ different.} \nonumber
\end{eqnarray}
The number of times $t_j$ that $X_j$ appears in each row and column of the matrix is obtained as follows:
\begin{eqnarray}
t_j = \prod_{s \in S_j} (a_s - 1), \nonumber
\end{eqnarray}
where $j = 0,1,2,\dots,2^k-1$, and $S_j$ is the set of indices of attributes that differ between input and output data in the event associated with $X_j$. Note that we define $t_0 = 1$.

First, we discuss the inequality relations among the elements from $X_0$ to $X_{2^k-1}$. Given that it is expected to output data closer to the true data with a higher probability in the randomized response, the following inequalities should hold (as in the case of three-attribute data):
\begin{eqnarray}
&& X_0 \geq X_1,\ X_0 \geq X_2,\ \dots,\ X_0 \geq X_k, \nonumber\\
&& X_1 \geq X_j\ (\mathrm{If}\ |S_j| = 2 \land S_1 \subset S_j), \nonumber\\
&& X_2 \geq X_j\ (\mathrm{If}\ |S_j| = 2 \land S_2 \subset S_j), \nonumber\\
&& \ \ \ \ \ \vdots \nonumber\\
&& X_k \geq X_j\ (\mathrm{If}\ |S_j| = 2 \land S_k \subset S_j), \nonumber\\
&& X_{1+k} \geq X_j\ (\mathrm{If}\ |S_j| = 3 \land S_{1+k} \subset S_j), \nonumber\\
&& \ \ \ \ \ \vdots \nonumber\\
&& X_{k+{}_k\mathrm{C}_{2}} \geq X_j\ (\mathrm{If}\ |S_j| = 3 \land S_{k+{}_k\mathrm{C}_{2}} \subset S_j), \nonumber\\
&& \ \ \ \ \ \vdots \nonumber\\
&& X_{1+{}_k\mathrm{C}_{1}+\cdots+{}_k\mathrm{C}_{k-3}} \geq X_j\ \nonumber\\
&& \ \ \ \ (\mathrm{If}\ |S_j| = k-1 \land S_{1+{}_k\mathrm{C}_{1}+\cdots+{}_k\mathrm{C}_{k-3}} \subset S_j), \nonumber\\
&& \ \ \ \ \ \vdots \nonumber\\
&& X_{{}_k\mathrm{C}_{1}+\cdots+{}_k\mathrm{C}_{k-2}} \geq X_j\ \nonumber\\
&& \ \ \ \ (\mathrm{If}\ |S_j| = k-1 \land S_{{}_k\mathrm{C}_{1}+\cdots+{}_k\mathrm{C}_{k-2}} \subset S_j), \nonumber\\
&& X_{1+{}_k\mathrm{C}_{1}+\cdots+{}_k\mathrm{C}_{k-2}} \geq X_{2^k-1}, \nonumber\\
&& \ \ \ \ \ \vdots \nonumber\\
&& X_{{}_k\mathrm{C}_{1}+\cdots+{}_k\mathrm{C}_{k-1}} \geq X_{2^k-1}. \nonumber
\end{eqnarray}
Therefore, we consider a linear programming problem of minimizing the value of $x_0 = X_0 / X_{2^k-1}$, which indicates the privacy level for the entire data satisfied from $\mathbf{P}$. Similar to $x_0$, we let $x_j = X_j/X_{2^k-1}$ for $j = 1,2,\dots, 2^k-1$. Here, the variables for the problem are from $x_0$ to $x_{2^k-2}$ because $x_{2^k-1} = 1$, and the number of inequality relations is $\sum_{i=0}^{k-1} (k-i) \cdot {}_{k}\mathrm{C}_{i} = k \cdot 2^{k-1}$. Then, we can construct $\mathbf{A}_{ub}$ and $\mathbf{b}_{ub}$, where the matrix and vector sizes are $k \cdot 2^{k-1} \times 2^{k}-1$ and $k \cdot 2^{k-1}$, respectively.

Next, we discuss the equality relations regarding the privacy levels of $k$ attributes, $\epsilon_1$, $\epsilon_2$, $\dots$, $\epsilon_k$. Here, we consider the relations for $\epsilon_i$ ($i \in \{1,2,\dots,k\}$). Because
\begin{eqnarray}
\hspace{-0.5cm} && \Pr[\mathrm{Output}[i] = u | \mathrm{Input}[i] = u] = \sum_{j\, :\, i \notin S_j} X_j \cdot t_j \nonumber\\
\hspace{-0.5cm} &\mathrm{and}& \Pr[\mathrm{Output}[i] = u | \mathrm{Input}[i] = v] = \sum_{j\, :\, i \in S_j} \frac{X_j \cdot t_j}{a_i - 1} \nonumber
\end{eqnarray}
where $v \neq u$, the relation (2) can be rewritten as
\begin{eqnarray}
\frac{x_0 + \sum_{j\, :\, j \in \{1,\dots,2^k-2\} \land i \notin S_j} x_j \cdot t_j}{\sum_{j\, :\, j \in \{1,\dots,2^k-2\} \land i \in S_j} \frac{x_j \cdot t_j}{a_i - 1} + \frac{\prod_{j=1}^k (a_j - 1)}{a_i-1}} = e^{\epsilon_i} 
\end{eqnarray}
using $t_0 = 1$, $S_{0} = \emptyset$, $x_{2^k-1} = 1$, and $S_{2^k-1} = \{1,2,\dots,k\}$. Then, we can construct $\mathbf{A}_{eq}$ and $\mathbf{b}_{eq}$, where the matrix and vector sizes are $k \times 2^k-1$ and $k$, respectively.

Based on the above discussion, we summarize the procedure for obtaining $\mathbf{A}_{ub}$, $\mathbf{b}_{ub}$, $\mathbf{A}_{eq}$, and $\mathbf{b}_{eq}$ in Algorithms 1 and 2. As $\mathbf{A}_{ub}$ and $\mathbf{b}_{ub}$ can be obtained from only the number of attributes $k$, it would be preferable to prepare fixed ones regardless of the dataset. $\mathbf{A}_{eq}$ and $\mathbf{b}_{eq}$ can be easily obtained from the equation (3), although the number of possible attribute values and the privacy levels are required.
\vspace{-0.2cm}
\begin{algorithm}
 \caption{Algorithm for constructing $\mathbf{A}_{ub}$ and $\mathbf{b}_{ub}$.}
 \begin{algorithmic}[1]
 \renewcommand{\algorithmicrequire}{\textbf{Input:}}
 \renewcommand{\algorithmicensure}{\textbf{Output:}}
 \REQUIRE The number of attributes $k$.
 \ENSURE $\mathbf{A}_{ub}$, $\mathbf{b}_{ub}$\\
 \STATE Let $l = [1,2,\dots,k]$.
 \STATE Make a list $comb$ such that $comb[h]$ represents the set of sets containing $h$ elements in $l$.
 \STATE Consider $S_j$ based on $comb$ and compute $t[j] = \prod_{s \in S_j} (a_s - 1)$ for $j = 0,1,\dots,2^k-1$.
 \STATE Construct an integer vector $p$ such that $p[h]$ represents the minimum of $j$ that satisfies $|S_j| = h+1$; that is, $p[h] = 1 + {}_{k}\mathrm{C}_{1} + \cdots + {}_{k}\mathrm{C}_{h}$.\\
 \ \\
 {\it \# Construction of $\mathbf{A}_{ub} \in \mathbb{R}^{k\cdot2^{k-1} \times 2^{k}-1}$}
 \STATE Set $r = 0$ and $g = 1$.
 \STATE Create a matrix $\mathbf{A}_{ub}$ whose all elements are $0$.
 \FOR{$h \in \{1,2,\dots,k\}$} 
 \STATE $\mathbf{A}_{ub}[r][0] = -1$; $\mathbf{A}_{ub}[r][h] = 1$
 \STATE $r \leftarrow r + 1$
 \ENDFOR
 \FOR{$h \in \{1,2,\dots,k-2\}$}
 \FOR{$i \in \{0,1,\dots,{}_{k}\mathrm{C}_{h}-1\}$}
 \FOR{$j \in \{0,1,\dots,{}_{k}\mathrm{C}_{h+1}-1\}$}
 \STATE Compute \begin{eqnarray}
 f = \begin{cases}
     1 \ \ \ (\mathrm{if}\ comb[h][i] \subset comb[h+1][j])\\
     0 \ \ \ \mathrm{(otherwise)}
 \end{cases}. \nonumber
 \end{eqnarray}
 \IF{$f = 1$}
 \STATE $\mathbf{A}_{ub}[r][g] = -1$; $\mathbf{A}_{ub}[r][p[h]+j] = 1$
 \STATE $r \leftarrow r + 1$
 \ENDIF
 \ENDFOR
 \STATE $g \leftarrow g + 1$
 \ENDFOR
 \ENDFOR
 \FOR{$h \in \{0,1,\dots,k-1\}$}
 \STATE $\mathbf{A}_{ub}[r+h][p[k-2]+h] = -1$
 \ENDFOR\\
 \ \\
 {\it \# Construction of $\mathbf{b}_{ub} \in \mathbb{R}^{k \cdot 2^{k-1}}$}
 \STATE Create a vector $\mathbf{b}_{ub}$ whose all elements are $0$.
 \FOR{$h \in \{1,\dots,k\}$}
 \STATE $\mathbf{b}_{ub}[k \cdot 2^{k-1} - h] = -1$
 \ENDFOR\\
 \ \\
 \RETURN $\mathbf{A}_{ub}$, $\mathbf{b}_{ub}$
 \end{algorithmic} 
\end{algorithm}

\begin{algorithm}
 \caption{Algorithm for constructing $\mathbf{A}_{eq}$ and $\mathbf{b}_{eq}$.}
 \begin{algorithmic}[1]
 \renewcommand{\algorithmicrequire}{\textbf{Input:}}
 \renewcommand{\algorithmicensure}{\textbf{Output:}}
 \REQUIRE The numbers of possible values $a_1, a_2, \dots, a_k$ and the privacy levels $\epsilon_1, \epsilon_2, \dots, \epsilon_k$ for $k$-attribute data.
 \ENSURE $\mathbf{A}_{eq}$, $\mathbf{b}_{eq}$\\
 \STATE Perform the same process as Steps 1 -- 4 in Algorithm 1.\\
 \ \\
 {\it \# Construction of $\mathbf{A}_{eq} \in \mathbb{R}^{k \times 2^{k}-1}$}
 \FOR{$h \in \{1,2,\dots,k\}$}
 \STATE $\mathbf{A}_{eq}[h-1][0] = 1$
 \STATE Set $c = 1$.
 \FOR{$i \in \{1,2,\dots,k-1\}$}
 \FOR{$j \in \{0,1,\dots,{}_{k}\mathrm{C}_{i}-1\}$}
 \STATE Compute \begin{eqnarray}
 f = \begin{cases}
     1 \ \ \ (\mathrm{if}\ h \in comb[i][j])\\
     0 \ \ \ \mathrm{(otherwise)}
 \end{cases}. \nonumber
 \end{eqnarray}
 \IF{$f = 1$}
 \STATE $\mathbf{A}_{eq}[h-1][c] = - (e^{\epsilon_h} \cdot t[c])/(a_h - 1)$
 \ELSE
 \STATE $\mathbf{A}_{eq}[h-1][c] = t[c]$
 \ENDIF
 \STATE $c \leftarrow c + 1$
 \ENDFOR
 \ENDFOR
 \ENDFOR\\
 \ \\
 {\it \# Construction of $\mathbf{b}_{eq} \in \mathbb{R}^{k}$}
 \FOR{$h \in {1,2,\dots,k}$}
 \STATE $\mathbf{b}_{eq}[h-1] = \left( e^{\epsilon_h} \cdot \prod_{i = 1}^k (a_i - 1) \right) / (a_h - 1)$
 \ENDFOR\\
 \ \\
 \RETURN $\mathbf{A}_{eq}$, $\mathbf{b}_{eq}$
 \end{algorithmic} 
\end{algorithm}
\vspace{-0.1cm}
Using the obtained $\mathbf{A}_{ub}$, $\mathbf{b}_{ub}$, $\mathbf{A}_{eq}$, and $\mathbf{b}_{eq}$, the linear programming problem of finding the optimal distortion matrix for $k$-attribute data, as in the three-attribute case, can be formulated as follows:
\begin{eqnarray}
\min_{\mathbf{x}} && (1,0,0,\dots,0) \cdot \mathbf{x} \nonumber\\
\mathrm{s.t.} && \mathbf{A}_{ub} \cdot \mathbf{x} \leq \mathbf{b}_{ub}, \nonumber\\
&& \mathbf{A}_{eq} \cdot \mathbf{x} = \mathbf{b}_{eq}. \nonumber
\end{eqnarray}

The distortion matrix constructed by solving this problem can provide the privacy-optimized randomized response mechanism that achieves the strongest privacy guarantees for the entire dataset while maintaining the privacy level of each attribute information.

However, a slight increase in $k$ makes the number of variables $2^k-1$ extremely large, and when $k$ is over $10$, finding an optimal solution in a practical time is difficult. Therefore, in the next subsection, we propose an efficient method for providing a near-optimal solution even for larger datasets with $100$ or more attributes. 
\vspace{-0.1cm}
\subsection{Heuristic Method}

We propose an efficient heuristic method that can be performed in $\mathcal{O}(k^2)$ time for $k$-attribute data. In our method, we consider that the probabilities are the same for all cases where two or more attribute values differ between input and output data. Specifically, among the variables $x_0, x_1, \dots, x_{2^k-1}$ described in Section IV.A.2, the values of $x_{k+1}, x_{k+2}, \dots, x_{2^k-1}$ are set to $1$. Then, $k+1$ variables from $x_0$ to $x_k$ can be inductively computed based on the case of two-attribute data. 

In the following, we first present the optimal solution for two-attribute data where the possible values for each attribute information are $m$ and $n$, along with its proof. Thereafter, we describe using it to find a near-optimal solution for three-attribute data. In addition, we generalize the procedure and show an inductive method for obtaining a near-optimal solution for $k$-attribute data from the solution for $(k-1)$-attribute data in $\mathcal{O}(k)$ time. Overall, the time complexity to construct a near privacy-optimized randomized response for $k$-attribute data is $\mathcal{O}(k^2)$.

\subsubsection{Optimal Mechanism for $m \times n$ Data}

Let $\epsilon_1$ and $\epsilon_2$ be the privacy level for each attribute information, respectively. There are four values that appear in the distortion matrix for randomized response: $X_0$, $X_1$, $X_2$, and $X_3$, and the inequality relations $X_0 \geq X_1, X_2 \geq X_3$ hold as in the discussion in Section IV.A. Here, we define
\begin{eqnarray}
x_{2,0} = \frac{X_0}{X_3},\ x_{2,1} = \frac{X_1}{X_3},\ \mathrm{and}\ \, x_{2,2} = \frac{X_2}{X_3}. \nonumber
\end{eqnarray}
Then, we can consider the values of $(x_{2,0}, x_{2,1}, x_{2,2})$ such that $x_{2,0}$ is minimized while satisfying the conditions of privacy levels and inequality relations. The solution of this problem for the case of $m \times n$ data is as follows:
\begin{eqnarray}
\hspace{-0.5cm} && (x_{2,0}, x_{2,1}, x_{2,2}) = \nonumber\\
\hspace{-0.5cm} && \begin{cases}
    \vspace{-0.3cm}
    \left( \frac{n \cdot e^{\epsilon_1 + \epsilon_2} + (m-1)(n-1)(e^{\epsilon_2} - 1)}{e^{\epsilon_2} + n - 1}, 1, \frac{n \cdot e^{\epsilon_1} - (m-1)(e^{\epsilon_2} - 1)}{e^{\epsilon_2} + n - 1} \right) \\
    \ \\
    \vspace{-0.3cm}
    \left( \frac{m \cdot e^{\epsilon_1 + \epsilon_2} + (m-1)(n-1)(e^{\epsilon_1} - 1)}{e^{\epsilon_1} + m - 1}, \frac{m \cdot e^{\epsilon_2} - (n-1)(e^{\epsilon_1} - 1)}{e^{\epsilon_1} + m - 1}, 1 \right)\\
    \ \\
    \vspace{-0.3cm}
    \left( \frac{(n-1)(e^{\epsilon_1} + m - 1)e^{\epsilon_2}}{-(e^{\epsilon_1} - 1) e^{\epsilon_2} + m(n-1)}, x_{2,0}, \frac{m(n-1)e^{\epsilon_1} + (m-1)(e^{\epsilon_1}-1)e^{\epsilon_2}}{-(e^{\epsilon_1} - 1) e^{\epsilon_2} + m(n-1)} \right)\\
    \ \\
    \left( \frac{(m-1)e^{\epsilon_1}(e^{\epsilon_2}+n-1)}{-e^{\epsilon_1} (e^{\epsilon_2} - 1) + (m-1)n}, \frac{(m-1)n e^{\epsilon_2} + (n-1)e^{\epsilon_1}(e^{\epsilon_2}-1)}{-e^{\epsilon_1} (e^{\epsilon_2} - 1) + (m-1)n}, x_{2,0} \right)
\end{cases} \hspace{-0.4cm} , \nonumber
\end{eqnarray}
where the cases are
\begin{eqnarray}
\hspace{-0.45cm}&\mathrm{(I)}& \hspace{-0.2cm} e^{\epsilon_1 + \epsilon_2} \geq (m-1)(n-1) \ \land \ n(e^{\epsilon_1}-1) \geq m(e^{\epsilon_2} - 1), \nonumber\\
\hspace{-0.45cm}&\mathrm{(II)}& \hspace{-0.2cm} e^{\epsilon_1 + \epsilon_2} \geq (m-1)(n-1) \ \land \ n(e^{\epsilon_1}-1) < m(e^{\epsilon_2} - 1), \nonumber\\
\hspace{-0.45cm}&\mathrm{(III)}& \hspace{-0.2cm} e^{\epsilon_1 + \epsilon_2} < (m-1)(n-1) \ \land \ \nonumber\\
\hspace{-0.45cm}&& (n-m)e^{\epsilon_1 + \epsilon_2} - m(n-1)e^{\epsilon_1} + (m-1)ne^{\epsilon_2} \geq 0, \nonumber\\
\hspace{-0.45cm}&\mathrm{(IV)}& \hspace{-0.2cm} e^{\epsilon_1 + \epsilon_2} < (m-1)(n-1) \ \land \ \nonumber\\
\hspace{-0.45cm}&& (n-m)e^{\epsilon_1 + \epsilon_2} - m(n-1)e^{\epsilon_1} + (m-1)ne^{\epsilon_2} < 0, \nonumber
\end{eqnarray}
in that order. 

We provide below the proof that this solution is optimal; that is, the obtained value of $x_{2,0}$ is the minimum under 
\begin{eqnarray}
&& \hspace{-0.3cm} \begin{cases}
\vspace{-0.3cm}
\frac{x_{2,0} + (n-1) \cdot x_{2,2}}{x_{2,1} + (n-1)} = e^{\epsilon_1}\\
\ \\
\frac{x_{2,0} + (m-1) \cdot x_{2,1}}{x_{2,2} + (m-1)} = e^{\epsilon_2}
\end{cases}\\
&\mathrm{and}& x_{2,0} \geq x_{2,1}, x_{2,2} \geq 1. \nonumber 
\end{eqnarray}

\begin{proof}
\begin{eqnarray}
(4) \hspace{-0.3cm} &\iff& \hspace{-0.3cm} \begin{cases}
x_{2,0} + (n-1) \cdot x_{2,2} = e^{\epsilon_1} \cdot x_{2,1} + (n-1) \cdot e^{\epsilon_1}\\
x_{2,0} + (m-1) \cdot x_{2,1} = e^{\epsilon_2} \cdot x_{2,2} + (m-1) \cdot e^{\epsilon_2}
\end{cases} \nonumber\\
\hspace{-0.3cm} &\iff& \hspace{-0.3cm} \begin{cases}
x_{2,1} = \frac{e^{\epsilon_2} + n-1}{e^{\epsilon_1} + m-1} \cdot x_{2,2} + \frac{(m-1) \cdot e^{\epsilon_2} - (n-1) \cdot e^{\epsilon_1}}{e^{\epsilon_1} + m-1}\\
x_{2,1} = \frac{e^{\epsilon_2}}{m-1} \cdot x_{2,2} + e^{\epsilon_2} - \frac{x_{2,0}}{m-1}
\end{cases}. \nonumber
\end{eqnarray}

When $\frac{e^{\epsilon_2}}{m-1} \geq \frac{e^{\epsilon_2} + n-1}{e^{\epsilon_1} + m-1} \iff e^{\epsilon_1 + \epsilon_2} \geq (m-1)(n-1)$, the values of $x_{2,1}$ and $x_{2,2}$ decrease as $x_{2,0}$ decreases. Because $x_{2,1}, x_{2,2} \geq 1$, $x_{2,0}$ is minimized when $x_{2,1} = 1$ or $x_{2,2} = 1$. When $x_{2,1} = 1$, 
\begin{eqnarray}
x_{2,2} = \frac{n \cdot e^{\epsilon_1} - (m-1)(e^{\epsilon_2} - 1)}{e^{\epsilon_2} + n - 1}. \nonumber
\end{eqnarray}
Therefore, when 
\begin{equation}
\frac{n \cdot e^{\epsilon_1} - (m-1)(e^{\epsilon_2} - 1)}{e^{\epsilon_2} + n - 1} \geq 1 \iff n(e^{\epsilon_1}-1) \geq m(e^{\epsilon_2} - 1), \nonumber
\end{equation}
we can obtain the optimal solution for case (I). $x_{2,0} \geq x_{2,2}$ also holds because
\begin{eqnarray}
x_{2,0} - x_{2,2} = \frac{n(e^{\epsilon_2}-1)(e^{\epsilon_1} + m-1)}{e^{\epsilon_2} + n-1} \geq 0. \nonumber
\end{eqnarray}
When $n(e^{\epsilon_1}-1) < m(e^{\epsilon_2} - 1)$, $x_{2,0}$ is minimized when $x_{2,2} = 1$, and we can obtain the solution for case (II). As in the above case, $x_{2,0} \geq x_{2,1}$ also certainly holds.

When $e^{\epsilon_1 + \epsilon_2} < (m-1)(n-1)$, $x_{2,1}$ and $x_{2,2}$ increase as $x_{2,0}$ decreases. Because $x_{2,1}, x_{2,2} \leq x_{2,0}$, $x_{2,0}$ is minimized when $x_{2,1} = x_{2,0}$ or $x_{2,2} = x_{2,0}$. When $x_{2,1} = x_{2,0}$, $(x_{2,0}, x_{2,2})$ is
{\footnotesize
\begin{eqnarray}
\ \left( \frac{(n-1)(e^{\epsilon_1} + m - 1)e^{\epsilon_2}}{-e^{\epsilon_1 + \epsilon_2} + e^{\epsilon_2} + m(n-1)}, \frac{m(n-1)e^{\epsilon_1} + (m-1)(e^{\epsilon_1}-1)e^{\epsilon_2}}{-e^{\epsilon_1 + \epsilon_2} + e^{\epsilon_2} + m(n-1)} \right). \nonumber
\end{eqnarray}}
When $x_{2,2} = x_{2,0}$, $(x_{2,0}, x_{2,1})$ is
{\footnotesize
\begin{eqnarray}
\ \left( \frac{(m-1)e^{\epsilon_1}(e^{\epsilon_2}+n-1)}{-e^{\epsilon_1 + \epsilon_2} + e^{\epsilon_1} + (m-1)n}, \frac{(m-1)n e^{\epsilon_2} + (n-1)e^{\epsilon_1}(e^{\epsilon_2}-1)}{-e^{\epsilon_1 + \epsilon_2} + e^{\epsilon_1} + (m-1)n} \right). \nonumber
\end{eqnarray}}
\hspace{-0.21cm} Given $x_{2,0} \geq x_{2,1}, x_{2,2}$, we can obtain the optimal solutions for cases (III) and (IV). In the both cases, $x_{2,1}, x_{2,2} \geq 1$ certainly holds as in cases (I) and (II).
\end{proof}

Using this optimal solution as a basis, we can obtain near-optimal solutions for the cases of $3$, $4$, $\dots$, $k$-attribute inductively. In the following, we first show how to compute the solution for three-attribute data using $x_{2,0}$, $x_{2,1}$, and $x_{2,2}$, and then present an inductive method for obtaining the solution for $k$-attribute data from the solution for $(k-1)$-attribute data.

\subsubsection{Inductive Method for Three-Attribute Data}

Let $\epsilon_1$, $\epsilon_2$, and $\epsilon_3$ be the privacy levels and $a_1$, $a_2$, and $a_3$ be the numbers of possible attribute values. Here, we consider the following five values that appear in the distortion matrix (as in Section IV.A):
\begin{eqnarray}
\hspace{-0.5cm} && X_{3,0} \ : \ \mathrm{Input\ and\ output\ data\ are\ exactly\ the\ same}; \nonumber\\
\hspace{-0.5cm} && X_{3,1} \ : \ \mathrm{Only\ the\ first\ attrubute\ value\ differs}; \nonumber\\
\hspace{-0.5cm} && X_{3,2} \ : \ \mathrm{Only\ the\ second\ attrubute\ value\ differs}; \nonumber\\
\hspace{-0.5cm} && X_{3,3} \ : \ \mathrm{Only\ the\ third\ attrubute\ value\ differs}; \nonumber\\
\hspace{-0.5cm} && X_{3,4} \ : \ \mathrm{More\ than\ one\ attribute\ values\ differ}, \nonumber
\end{eqnarray}
by assuming all probabilities of events that hold the condition where two or more attribute values differ to be equal; that is, $X_4 = X_5 = X_6 = X_7$ in Section IV.A.1. Then, there are four variables to consider:
\begin{eqnarray}
\ \ x_{3,0} = \frac{X_{3,0}}{X_{3,4}}, \ x_{3,1} = \frac{X_{3,1}}{X_{3,4}}, \ x_{3,2} = \frac{X_{3,2}}{X_{3,4}}, \ x_{3,3} = \frac{X_{3,3}}{X_{3,4}}, \nonumber
\end{eqnarray}
and from (2), the values of $e^{\epsilon_1}$, $e^{\epsilon_2}$, and $e^{\epsilon_3}$ are calculated as
{\footnotesize
\begin{eqnarray}
\frac{x_{3,0} + (a_2 - 1) x_{3,2} + (a_3 - 1) x_{3,3} + a_2 a_3 - (1 + a_2 - 1 + a_3 - 1)}{x_{3,1} + a_2 a_3 -1}, \nonumber\\
\ \nonumber\\
\vspace{-0.3cm}
\frac{x_{3,0} + (a_1 - 1) x_{3,1} + (a_3 - 1) x_{3,3} + a_1 a_3 - (1 + a_1 - 1 + a_3 - 1)}{x_{3,2} + a_1 a_3 -1}, \nonumber\\
\ \nonumber\\
\vspace{-0.3cm}
\frac{x_{3,0} + (a_1 - 1) x_{3,1} + (a_2 - 1) x_{3,2} + a_1 a_2 - (1 + a_1 - 1 + a_2 - 1)}{x_{3,3} + a_1 a_2 -1}, \nonumber
\end{eqnarray}}
\hspace{-0.25cm} respectively. Among these, $e^{\epsilon_1}$ and $e^{\epsilon_2}$ can also be calculated using $x_{2,0}$, $x_{2,1}$, and $x_{2,2}$; therefore, the following relations should hold:
\begin{eqnarray}
\begin{cases}
\frac{x_{2,0} + (a_2 - 1) x_{2,2}}{x_{2,1} + a_2 - 1} \\
\vspace{-0.3cm} \ \ \ \ \ = \frac{x_{3,0} + (a_2 - 1) x_{3,2} + (a_3 - 1) x_{3,3} + a_2 a_3 - (1 + a_2 - 1 + a_3 - 1)}{x_{3,1} + a_2 a_3 -1}\\
\ \\
\frac{x_{2,0} + (a_1 - 1) x_{2,1}}{x_{2,2} + a_1 - 1} \\
\ \ \ \ \ = \frac{x_{3,0} + (a_1 - 1) x_{3,1} + (a_3 - 1) x_{3,3} + a_1 a_3 - (1 + a_1 - 1 + a_3 - 1)}{x_{3,2} + a_1 a_3 -1}
\end{cases} \hspace{-0.7cm} 
\end{eqnarray}
Here, if
\begin{eqnarray}
\begin{cases}
x_{3,0} + (a_3 - 1) x_{3,3} = a_3 \cdot x_{2,0} \\
x_{3,1} + a_3 - 1 = a_3 \cdot x_{2,1} \\
x_{3,2} + a_3 - 1 = a_3 \cdot x_{2,2}
\end{cases}, 
\end{eqnarray}
the relations in (5) hold. The proof is provided below.

\begin{proof}
When the relations in (6) hold, 
\begin{eqnarray}
\hspace{-0.5cm} && x_{3,0} + (a_2 - 1) x_{3,2} + (a_3 - 1) x_{3,3} \nonumber\\
\hspace{-0.5cm} && \ \ \ \ \ \ \ \ + a_2 a_3 - (1 + a_2 - 1 + a_3 - 1) \nonumber\\
\hspace{-0.5cm} &=&  a_3 \cdot x_{2,0} + (a_2 - 1) \left( a_3 \cdot x_{2,2} - (a_3 - 1) \right) \nonumber\\
\hspace{-0.5cm} && \ \ \ \ \ \ \ \ \ \ \ \ \ \ \ \ \ \ \ \ \ \ \ \ \ + a_2 a_3 - a_2 - a_3 + 1 \nonumber\\
\hspace{-0.5cm} &=& a_3 \cdot x_{2,0} + a_3 \cdot (a_2 - 1) x_{2,2} \nonumber\\
\hspace{-0.5cm} && \ \ \ \ \ \ \ \ - (a_2 - 1)(a_3 - 1) + (a_2 - 1)(a_3 - 1) \nonumber\\
\hspace{-0.5cm} &=& a_3 \left( x_{2,0} + (a_2 - 1) x_{2,2} \right) \nonumber
\end{eqnarray}
and
\begin{eqnarray}
x_{3,1} + a_2 a_3 - 1 &=& a_3 \cdot x_{2,1} - (a_3 - 1) + a_2 a_3 - 1 \nonumber\\
&=& a_3 (x_{2,1} + a_2 - 1). \nonumber
\end{eqnarray}
Therefore, for the first equality in (5), the right-hand side is 
\begin{eqnarray}
\frac{a_3 \left( x_{2,0} + (a_2 - 1) x_{2,2} \right)}{a_3 (x_{2,1} + a_2 - 1)} = \frac{x_{2,0} + (a_2 - 1) x_{2,2}}{x_{2,1} + a_2 - 1}, \nonumber
\end{eqnarray}
which is identical to the left-hand side. Similarly, for the second equality, the right-hand side equals the left-hand side when the relations in (6) hold.
\end{proof}

Therefore, for the values from $x_{3,0}$ to $x_{3,3}$ to satisfy the equality relations for privacy levels, we first obtain $x_{3,1}$ and $x_{3,2}$ as
\begin{eqnarray}
\hspace{-0.5cm} && x_{3,1} = a_3 \cdot x_{2,1} - a_3 + 1 \nonumber\\
\hspace{-0.5cm} &\mathrm{and}& x_{3,2} = a_3 \cdot x_{2,2} - a_3 + 1 \nonumber
\end{eqnarray}
from the second and third equality relations in (6). Thereafter, using these values, we can solve the following simultaneous linear equations:
\begin{eqnarray}
\begin{cases}
\vspace{-0.3cm}
x_{3,0} + (a_3 - 1)x_{3,3} = a_3 \cdot x_{2,0} \\
\ \\
\frac{x_{3,0} + (a_1 - 1) x_{3,1} + (a_2 - 1) x_{3,2} + a_1 a_2 - (1 + a_1 - 1 + a_2 - 1)}{x_{3,3} + a_1 a_2 -1} = e^{\epsilon_3}
\end{cases} \nonumber
\end{eqnarray}
and obtain the values of $x_{3,0}$ and $x_{3,3}$.

From the above procedure, all values from $x_{3,0}$ to $x_{3,3}$ can be derived using the values from $x_{2,0}$ to $x_{2,2}$, the optimal solution for the two-attribute case. By constructing a distortion matrix from these values, a randomized response mechanism for three-attribute data that satisfies a near-optimal privacy guarantee for the entire dataset while maintaining the privacy level for each attribute information is expected to be provided.

\subsubsection{Inductive Method for $k$-Attribute Data}

By generalizing the method for three-attribute data, we propose an inductive method for obtaining a near-optimal solution for $k$-attribute data using the solution for $(k-1)$-attribute data.

Let $\epsilon_1, \epsilon_2, \dots, \epsilon_k$ be the privacy levels and $a_1, a_2, \dots, a_k$ be the numbers of possible attribute values for the $k$ attributes. The variables to consider are
\begin{eqnarray}
\ \ x_{k,0} = \frac{X_{k,0}}{X_{k,k+1}},\ x_{k,1} = \frac{X_{k,1}}{X_{k,k+1}},\ \dots,\ x_{k,k} = \frac{X_{k,k}}{X_{k,k+1}}, \nonumber
\end{eqnarray}
where $X_{k,j}$ $(j = 1,2,\dots,k)$ represents the probability of events that hold the condition where only the $j$-th attribute value differs between the input and output data, and the conditions associated with $X_{k,0}$ and $X_{k,k+1}$ are that the input and output data are exactly the same and that more than one attribute values differ, respectively. As mentioned in Section IV.A, the inequality relation $x_{k,0} \geq x_{k,1}, x_{k,2}, \dots, x_{k,k} \geq 1$ should hold. Here, we consider calculating the values of these $k+1$ variables using $x_{k-1,0}, x_{k-1,1}, \dots, x_{k-1,k-1}$ for the former $(k-1)$-attribute information.

Similar to the three-attribute case, the values of $e^{\epsilon_j}$ $(j = 1,2,\dots,k)$ are calculated as
\begin{eqnarray}
\frac{x_{k,0} + A(k,j) + B(k,j)}{x_{k,j} + C(k,j)}, \nonumber
\end{eqnarray}
where
\begin{eqnarray}
&& A(k,j) \ := \ \sum_{i=1}^{k} (a_i - 1) x_{k,i} - (a_j - 1) x_{k,j}, \nonumber \\
&& B(k,j) \ := \ \frac{\prod_{i=1}^k a_i}{a_j} - \left( \sum_{i=1}^k a_i - a_j - k + 2 \right), \nonumber\\
&\mathrm{and}& C(k,j) \ := \ \frac{\prod_{i = 1}^k a_i}{a_j} - 1. \nonumber
\end{eqnarray}
Specifically for $e^{\epsilon_k}$, the following equality holds:
\begin{eqnarray}
\hspace{-1.1cm} && \frac{x_{k,0} + \sum_{i=1}^{k-1} (a_i - 1) x_{k,i} + \prod_{i=1}^{k-1} a_i - \sum_{i=1}^{k-1} a_i + k - 2}{x_{k,k} + \prod_{i=1}^{k-1} a_i - 1} \nonumber\\
\hspace{-1.1cm} && \ \ \ \ \ \ \ \ \ \ \ \ \ \ \ \ \ \ \ \ \ \ \ \ \ \ \ \ \ \ \ \ \ \ \ \ \ \ \ \ \ \ \ \ \ \ \ \ \ \ \ \ \ \ \ \ \ \ = e^{\epsilon_k}. 
\end{eqnarray}
Here, $e^{\epsilon_{j}}$ $(j \in \{1,2,\dots,k-1\})$ can be calculated also using the values from $x_{k-1,0}$ to $x_{k-1,k-1}$. 
Therefore, for all $j \in \{1,2,\dots,k-1\}$, the following relation between $x_{k-1,0}, \dots, x_{k-1,k-1}$ and $x_{k,0}, \dots, x_{k,k}$ holds:
\begin{eqnarray}
\hspace{-0.5cm} && \hspace{-0.5cm} \frac{x_{k-1,0} + A(k-1,j) + B(k-1,j)}{x_{k-1,j} + C(k-1,j)} \nonumber\\
\hspace{-0.5cm} && \ \ \ \ \ \ \ \ \ \ \ \ \ \ \ \ \ \ \ = \frac{x_{k,0} + A(k,j) + B(k,j)}{x_{k,j} + C(k,j)}. 
\end{eqnarray}
Here, if
\begin{eqnarray}
\begin{cases}
x_{k,0} + (a_k - 1) x_{k,k} \ = \ a_k \cdot x_{k-1,0} \\
x_{k,j} + a_k - 1 \ = \ a_k \cdot x_{k-1, j} \ \ \ (j = 1,2,\dots,k-1)
\end{cases} \hspace{-0.3cm} , 
\end{eqnarray}
the relation (8) holds. The proof is provided below.

\begin{proof}
When the relations in (9) hold,
\begin{eqnarray}
&& x_{k,0} + A(k,j) + B(k,j) \nonumber\\
&=& x_{k,0} + (a_k - 1) x_{k,k} + \sum_{i=1}^{k-1} (a_i - 1) x_{k,i} - (a_j - 1) x_{k,j} \nonumber\\
&& + \frac{\prod_{i=1}^k a_i}{a_j} - \left( \sum_{i=1}^k a_i - a_j - k + 2 \right) \nonumber\\
&=& a_k \cdot x_{k-1,0} + \sum_{i=1}^{k-1} (a_i - 1) (a_k \cdot x_{k-1,i} - (a_k - 1)) \nonumber\\
&& - (a_j - 1) (a_k \cdot x_{k-1,j} - (a_k - 1)) \nonumber\\
&& + a_k \cdot \frac{\prod_{i=1}^{k-1} a_i}{a_j} - \left( a_k + \sum_{i=1}^{k-1} a_i \right) + a_j + k - 2 \nonumber\\
&=& a_k \cdot x_{k-1,0} + a_k \cdot \sum_{i=1}^{k-1} (a_i - 1) x_{k-1,i} - a_k \cdot \sum_{i=1}^{k-1} (a_i - 1)  \nonumber\\
&& + \sum_{i=1}^{k-1} (a_i - 1) - a_k \cdot (a_j - 1) x_{k-1,j} + (a_k - 1) (a_j - 1) \nonumber\\
&& + a_k \cdot \frac{\prod_{i=1}^{k-1} a_i}{a_j} - a_k - \sum_{i=1}^{k-1} a_i + a_j + k - 2 \nonumber\\
\hspace{-0.3cm} &=& a_k \cdot \Biggl( x_{k-1,0} + \sum_{i=1}^{k-1} (a_i - 1) x_{k-1,i} - (a_j - 1) x_{k-1,j} \nonumber\\
\hspace{-0.3cm} && \ \ \ \ \ \ \ \ \ \ + \frac{\prod_{i=1}^{k-1} a_i}{a_j} - \left( \sum_{i=1}^{k-1} a_i - a_j - k + 3 \right) \Biggr) \nonumber\\
\hspace{-0.3cm} &=& a_k \cdot \bigl( x_{k-1,0} + A(k-1,j) + B(k-1,j) \bigr)\nonumber
\end{eqnarray}
\vspace{-0.5cm}
and
\begin{eqnarray}
x_{k,j} + C(k,j) &=& a_k \cdot x_{k-1,j} - (a_k - 1) + \frac{\prod_{i=1}^k a_i}{a_j} - 1 \nonumber\\
&=& a_k \cdot \left( x_{k-1,j} + \frac{\prod_{i=1}^{k-1} a_i}{a_j} - 1 \right) \nonumber\\
&=& a_k \cdot \bigl( x_{k-1,j} + C(k-1,j) \bigr). \nonumber
\end{eqnarray}
Therefore, the right-hand side of (8) equals the left-hand side; that is, the relation (8) holds.
\end{proof}

Therefore, for the values from $x_{k,0}$ to $x_{k,k}$ to satisfy the equality relations regarding the privacy levels from $\epsilon_1$ to $\epsilon_{k}$, we first obtain $x_{k,j}$ $(j = 1,2,\dots,k-1)$ as
\begin{eqnarray}
x_{k,j} = a_k \cdot x_{k-1,j} - a_k + 1 \nonumber
\end{eqnarray}
from the second relation in (9). Then, using these values, we can solve the following simultaneous linear equations:
\begin{eqnarray}
\begin{cases}
x_{k,0} + (a_k - 1) x_{k,k} = a_k \cdot x_{k-1,0}\\
(7)
\end{cases} \nonumber
\end{eqnarray}
and obtain the values of $x_{k,0}$ and $x_{k,k}$.

From the above procedure, the values from $x_{k,0}$ to $x_{k,k}$ can be inductively derived from the values of $x_{k-1,0}$ to $x_{k-1,k-1}$. Here, it should be noted that the proposed heuristic method does not guarantee that $x_{k,0} \geq x_{k,j}$ for all $j \in \{1,2,\dots,k\}$. One possible way to deal with this is to recalculate the values of $x_{k,0}$ and $x_{k,k}$ as
\begin{eqnarray}
\begin{cases}
x_{k,0} = a_k \cdot x_{k-1,0} - a_k + 1 \\
x_{k,k} = 1
\end{cases} \nonumber
\end{eqnarray}
when $x_{k,0}$ or $x_{k,k}$ is less than $1$ or $x_{k,0} < x_{k,j}$ for some $j \in \{1,2,\dots, k\}$. This allows $x_{k,0} \geq x_{k,1}, \dots, x_{k,k} \geq 1$ when $x_{k-1,0} \geq x_{k-1,1}, \dots, x_{k-1,k-1} \geq 1$ because
\vspace{-0.1cm}
\begin{eqnarray}
\hspace{-0.55cm} && x_{k,0} = a_k \cdot x_{k-1,0} - a_k + 1 \geq a_k \cdot x_{k-1, j} - a_k + 1 = x_{k,j}, \nonumber\\
\hspace{-0.55cm} && x_{k,j} = a_k \cdot (x_{k-1,j} - 1) + 1 \geq 1 = x_{k,k} \nonumber
\end{eqnarray}
for $j = 1,2,\dots,k-1$. In this case, because the privacy level for the $k$-th attribute information will differ from the input, we should recompute $\epsilon_k$ using (7) and output as the privacy level that is actually guaranteed. \\

Overall, the heuristic method for constructing a near-optimal randomized response for $k$-attribute data can be summarized in Algorithm 3.

\begin{algorithm}
 \caption{Heuristic algorithm for inductively constructing a near privacy-optimized randomized response for $k$-attribute data.}
 \begin{algorithmic}[1]
 \renewcommand{\algorithmicrequire}{\textbf{Input:}}
 \renewcommand{\algorithmicensure}{\textbf{Output:}}
 \REQUIRE The number of possible values $a_1, a_2, \dots, a_k$ and desired privacy levels $\epsilon_1, \epsilon_2, \dots, \epsilon_k$ for $k$-attribute data.
 \ENSURE The values of $x_{k,0}, x_{k,1}, \dots, x_{k,k}$ for a near privacy-optimized randomized response and achieved privacy levels $\epsilon_1, \epsilon_2, \dots, \epsilon_k$.\\
 \ \\
 {\it \# Base Case}
 \STATE Compute the optimal values $(x_{2,0}, x_{2,1}, x_{2,2})$ for $a_1 \times a_2$ data using $\epsilon_1$ and $\epsilon_2$.\\
 \ \\
 {\it \# Induction Step}
 \FOR{$i \in \{3,4,\dots,k\}$}
 \FOR{$j \in \{1,2,\dots,i-1\}$}
 \STATE $x_{i,j} = a_i \cdot x_{i-1,j} - a_i + 1$
 \ENDFOR
 \STATE Compute $A(i,i) = \sum_{h=1}^{i-1} (a_h - 1) x_{i,h}$, $B(i,i) = \prod_{h=1}^{i-1} a_h - \sum_{h=1}^{i-1} a_h + i - 2$, and $C(i,i) = \prod_{h=1}^{i-1} a_h - 1$.
 \STATE Solve the following simultaneous linear equations:
 \begin{eqnarray}
 \begin{cases}
 x_{i,0} + (a_i - 1)x_{i,i} = a_{i} \cdot x_{i-1,0} \\
 x_{i,0} - e^{\epsilon_i} \cdot x_{i,i} = -A(i,i) - B(i,i) + e^{\epsilon_i} \cdot C(i,i)
 \end{cases} \nonumber
 \end{eqnarray}
 and obtain $x_{i,0}$ and $x_{i,i}$.
 \IF{$x_{i,0} < 1$ or $x_{i,i} < 1$ or $x_{i,0} < x_{i,j}$ for some $j \in \{1,2,\dots,i\}$}
 \STATE $x_{i,0} = a_i \cdot x_{i-1,0} - a_i + 1 \ \ \mathrm{and}\ \ x_{i,i} = 1$. 
 \STATE Using these values, recompute
 \begin{eqnarray}
 \epsilon_i = \ln\left({\frac{x_{i,0} + A(i,i) + B(i,i)}{x_{i,i} + C(i,i)}}\right). \nonumber
 \end{eqnarray}
 \ENDIF
 \ENDFOR\\
 \ \\
 \RETURN $x_{k,0}, x_{k,1}, \dots, x_{k,k}$ and $\epsilon_1, \epsilon_2, \dots, \epsilon_k$
 \end{algorithmic} 
\end{algorithm}

Step 1 for the base case can be computed in a constant time according to the discussion in Section IV.B.1, and the induction step can be performed in $\mathcal{O}(i)$ time for each $i$; therefore, the time complexity of Algorithm 3 is $\mathcal{O}(k^2)$. Thus, a near privacy-optimized randomized response mechanism can be obtained for large multi-attribute data of $k = 100$ or more in a practical time, which would be difficult with an approach that strictly solves the linear programming problem.

In the next section, we evaluate the optimality of our heuristic method and also show that it outperforms the existing Kronecker product-based method.

\section{Experiments and Discussion}

In the experiments, we first evaluated the strength of the guaranteed privacy offered by our two proposed methods for the entire dataset, and compared it with that of the existing Kronecker product-based method \cite{3}. Simultaneously, we investigated the effect of variation in the privacy level for each attribute information and the number of possible attribute values. In addition, we analyzed the optimality of our heuristic method by comparing it with our optimal solution from solving the linear programming problem. Furthermore, we provide an analysis example using our method and demonstrate its utility. Finally, we measured the run time of our methods and demonstrated that our heuristic method can be performed in a practical time even for large datasets with $k \geq 1,000$. The Python codes used in our experiments are available at \url{https://github.com/ay0408/Optimized-RR}.

\subsection{Privacy Level for the entire dataset}

We compared the achieved privacy levels for the entire dataset among our methods and the existing Kronecker product-based method, when the privacy level for each attribute information is given. 

\ \\
\vspace{-0.5cm}

\begin{figure}[htbt]
  (a) \ \ \ \ \ \ \ \ \ \ \ \ \ \ \ \ \ \ \ \ \ \ \ \ \ \ \ \ \ \ \ \ \ (b)\\
  \centerline{
  \includegraphics[width=4.5cm]{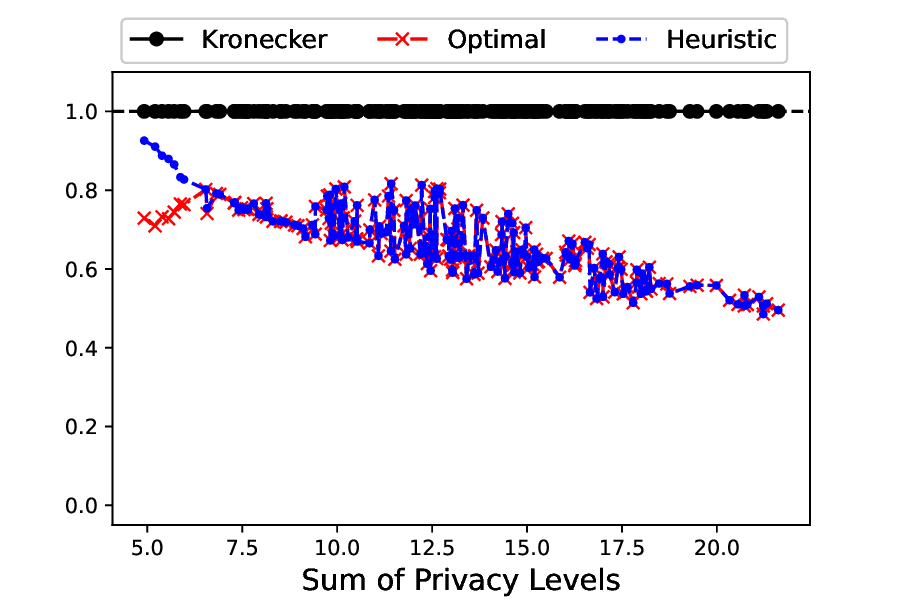}\ 
  \includegraphics[width=4.5cm]{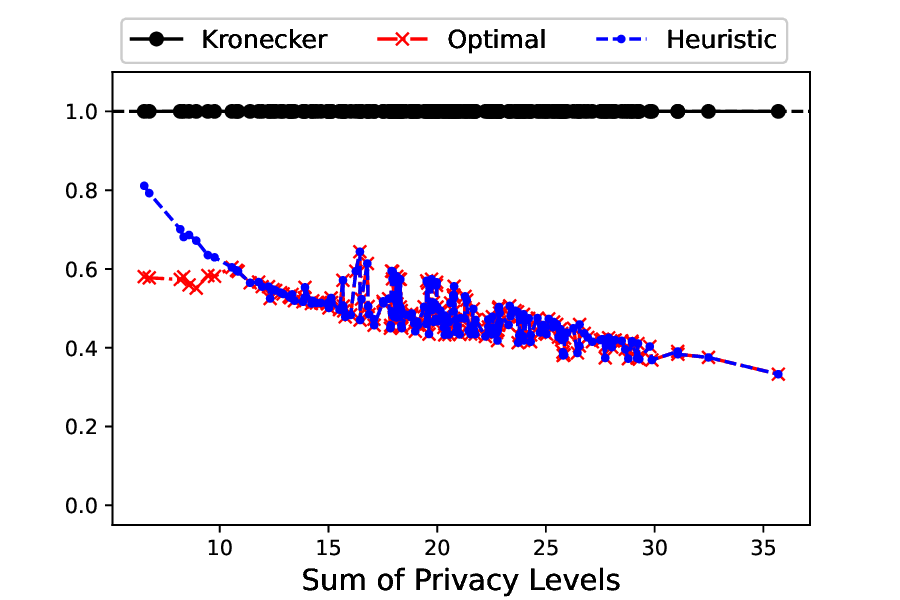}}
  
 (c) \ \ \ \ \ \ \ \ \ \ \ \ \ \ \ \ \ \ \ \ \ \ \ \ \ \ \ \ \ \ \ \ \ (d)\\
  \centerline{
  \includegraphics[width=4.5cm]{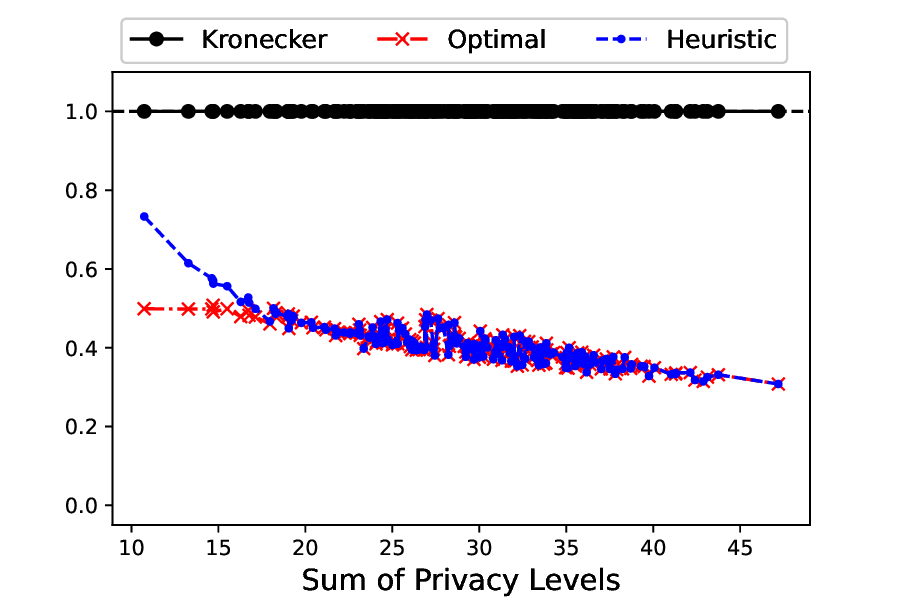}\ \includegraphics[width=4.5cm]{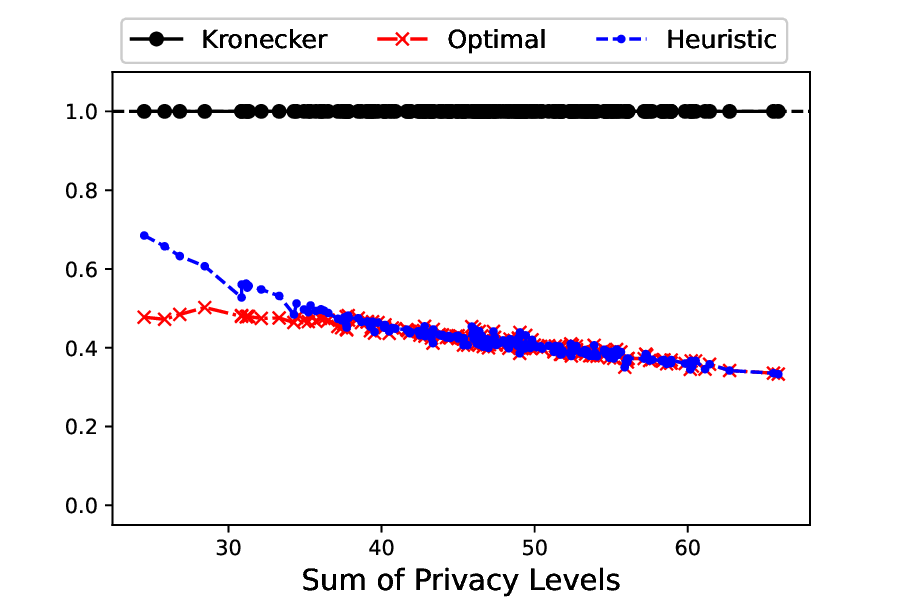}}
  \caption{Achieved privacy levels for the entire dataset when (a) $k = 3$, (b) $k = 5$, (c) $k = 7$, and (d) $k = 10$. The $x$-axis represents $\sum_{i=1}^k \epsilon_i$; that is, the sum of the privacy level for each attribute information. The $y$-axis represents the ratio of the achieved privacy level to $\sum_{i=1}^k \epsilon_i$. We compared the existing Kronecker product-based method (black, solid), the optimal mechanism (red, dash-dot), and our heuristic method (blue, dashed).}
  \label{fig1}
\end{figure}

First, we show in Fig. \ref{fig1} the results when we fixed $a_i = 5$ for all $i$. The value of each $\epsilon_i$ was randomly set from $1$ to $8$. For each $k \in \{3,5,7,10\}$, we generated $200$ different sets of $(\epsilon_1, \dots, \epsilon_k)$ and evaluated the performance of our methods on each set. In the figure, the results are sorted by $\sum_i^k \epsilon_i$, which also evaluates the impact of variation in the privacy level for each attribute information. Here, note that the privacy level achieved by the Kronecker product-based method is $\sum_{i=1}^k \epsilon_i$.

Fig. \ref{fig1} indicates that both of our methods provide significantly stronger privacy guarantees than the existing method regardless of the value of $k$. In particular, when $\sum_{i=1}^k \epsilon_i$ is relatively large, our heuristic method can construct a randomized response extremely close to the optimal. For a larger $k$, there is a narrower range of the achieved privacy levels for the same $\sum_{i=1}^k \epsilon_i$, which might be because the effect of the variation of each $\epsilon_i$ is less apparent. In addition, an important feature regarding the optimal solution is that when $\sum_{i=1}^k \epsilon_i$ is large, an increase in $\sum_{i=1}^k \epsilon_i$ provides much stronger privacy guarantees compared with the existing method, but the impact is not observed when $\sum_{i=1}^k \epsilon_i$ is small. In the following, we discuss this feature using the optimal solution for two-attribute data provided in Section IV.B.1. 

When $\sum_{i=1}^k \epsilon_i$ is large, we can compute the optimal value for cases (I) or (II). 
Here, we consider case (I). Because
\vspace{-0.1cm}
\begin{eqnarray}
\hspace{-0.3cm} \ln \left( \frac{n \cdot e^{\epsilon_1 + \epsilon_2} + (m-1)(n-1)(e^{\epsilon_2} - 1)}{e^{\epsilon_2} + n - 1} \right) \nonumber
\end{eqnarray}
\begin{eqnarray}
\hspace{-0.3cm} &=& \ln \left( n \cdot e^{\epsilon_1 + \epsilon_2} + (m-1)(n-1)(e^{\epsilon_2} - 1) \right) \nonumber\\
\hspace{-0.3cm} && \ \ \ \ \ \ \ \ \ \ \ \ \ \ \ \ \ \ \ \ \ \ \ \ \ \ \ \ \ \ \ \ \ \ - \ln \left( e^{\epsilon_2} + n - 1 \right) \nonumber\\
\hspace{-0.3cm} &=& \ln \left( n \cdot e^{\epsilon_1 + \epsilon_2} \right) + \ln \left( 1 + \frac{(m-1)(n-1)(e^{\epsilon_2} - 1)}{n \cdot e^{\epsilon_1 + \epsilon_2}} \right) \nonumber\\
\hspace{-0.3cm} && \ \ \ \ \ \ \ \ \ \ \ \ \ \ \ \ \ \ \ \ \ \ \ \ \ \ \ \ \ \ \ \ \ \ - \ln \left( e^{\epsilon_2} + n - 1 \right) \nonumber\\
\hspace{-0.3cm} &=& (\epsilon_1 + \epsilon_2) + \ln (n) + \ln \left( 1 + \frac{(m-1)(n-1)(e^{\epsilon_2} - 1)}{n \cdot e^{\epsilon_1 + \epsilon_2}} \right) \nonumber\\
\hspace{-0.3cm} && \ \ \ \ \ \ \ \ \ \ \ \ \ \ \ \ \ \ \ \ \ \ \ \ \ \ \ \ \ \ \ \ \ \  - \ln \left( e^{\epsilon_2} + n - 1 \right), 
\end{eqnarray}
when the values of $\epsilon_1$ and $\epsilon_2$ increase, the ratio of the optimal privacy level to $\epsilon_1 + \epsilon_2$ tends to decrease. We can make similar arguments for case (II).

When $\sum_i^k \epsilon_i$ is small, we can compute the optimal value for cases (III) or (IV). Here, we consider case (III) in a similar manner to the above discussion. The optimal privacy level for case (III) is
\begin{eqnarray}
\hspace{-0.5cm} && \ln \left( \frac{(n-1)(e^{\epsilon_1} + m - 1)e^{\epsilon_2}}{-(e^{\epsilon_1} - 1) e^{\epsilon_2} + m(n-1)} \right) \nonumber\\
\hspace{-0.5cm} &=&  (\epsilon_1 + \epsilon_2) + \ln (n-1) + \ln \left( 1 + \frac{(m-1) e^{\epsilon_2}}{e^{\epsilon_1 + \epsilon_2}} \right) \nonumber\\
\hspace{-0.5cm} && \ \ \ \ \ \ \ \ \ \ \ \ \ \ \ \ - \ln \left( - e^{\epsilon_1 + \epsilon_2} + e^{\epsilon_2} + m(n-1) \right). 
\end{eqnarray}
Unlike when $\sum_i^k \epsilon_i$ is large, no clear trend can be observed when the values of $\epsilon_1$ and $\epsilon_2$ vary. The same is true for case (IV).

\begin{figure}[htbt]
  (a) \ \ \ \ \ \ \ \ \ \ \ \ \ \ \ \ \ \ \ \ \ \ \ \ \ \ \ \ \ \ \ \ \ (b)\\
  \centerline{
  \includegraphics[width=4.5cm]{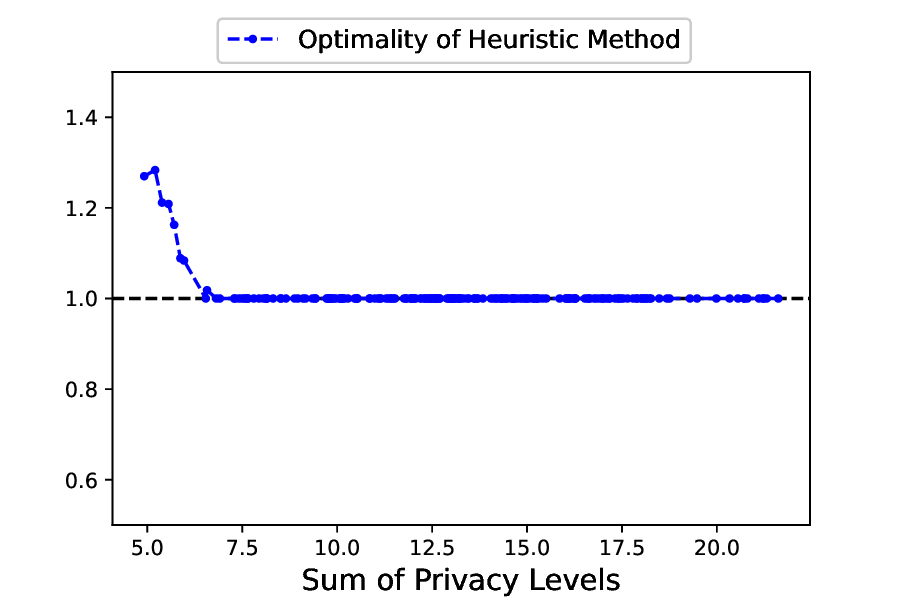}\ 
  \includegraphics[width=4.5cm]{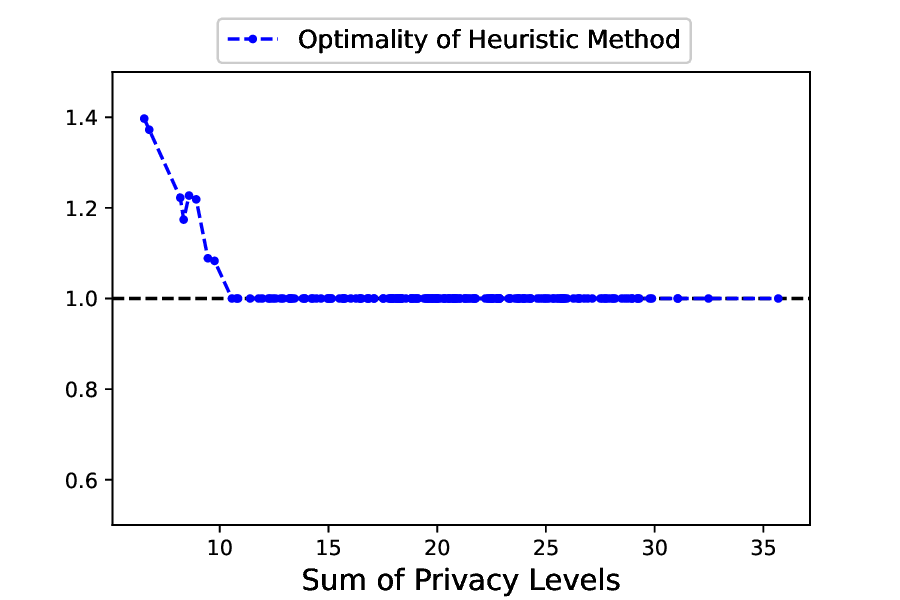}}
  
 (c) \ \ \ \ \ \ \ \ \ \ \ \ \ \ \ \ \ \ \ \ \ \ \ \ \ \ \ \ \ \ \ \ \ (d)\\
  \centerline{
  \includegraphics[width=4.5cm]{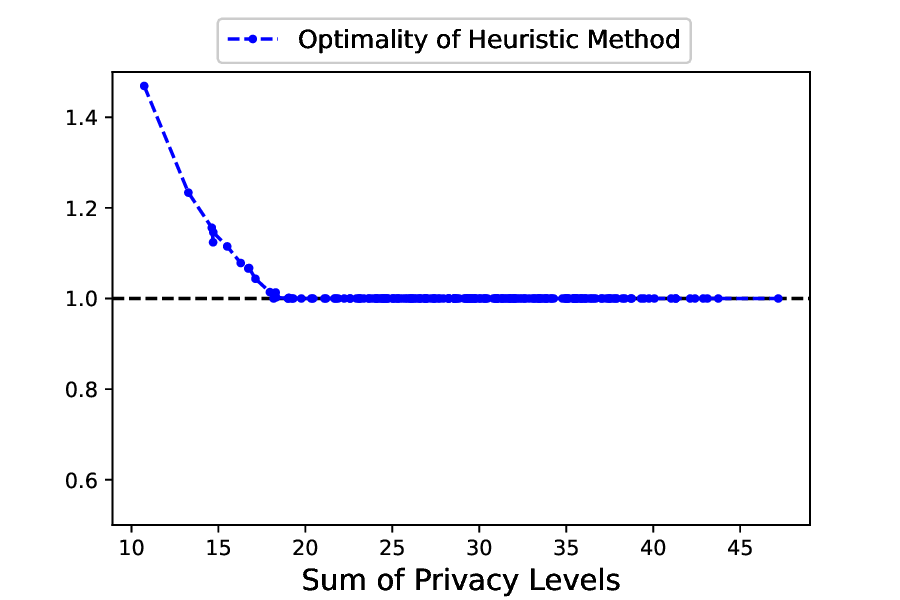}\ \includegraphics[width=4.5cm]{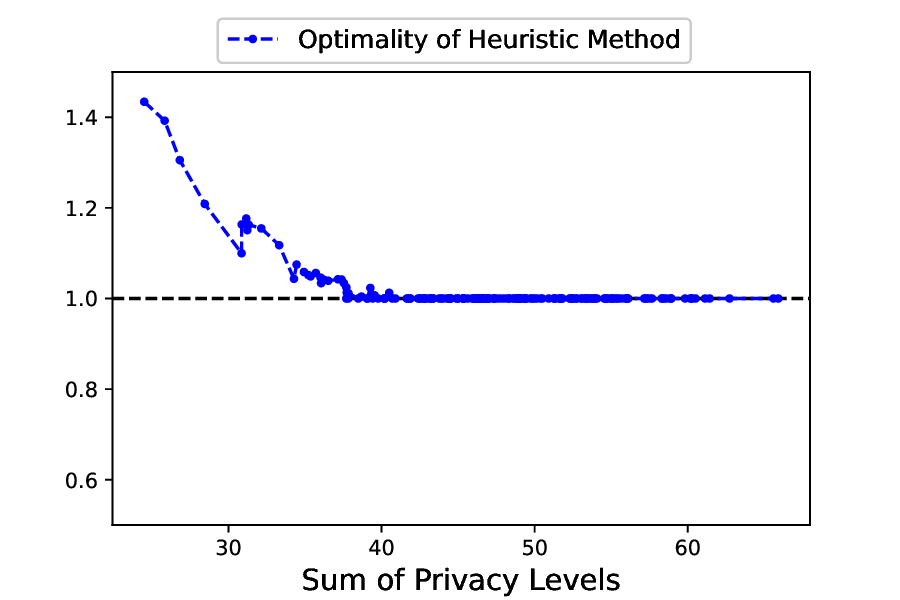}}
  \caption{Optimality of our heuristic method when (a) $k = 3$, (b) $k = 5$, (c) $k = 7$, and (d) $k = 10$. The $x$-axis represents $\sum_{i=1}^k \epsilon_i$; that is, the sum of the privacy level for each attribute information. The $y$-axis represents the ratio of the achieved privacy level to the optimal solution. }
  \label{fig11}
\end{figure}

While the above discussion is about the case of two-attribute data, we would expect the same feature to be observed for general $k$-attribute data, which appears in Fig. \ref{fig1}. Our heuristic method captures this characteristic of the optimal solution when $\sum_{i=1}^k \epsilon_i$ is large, but not when $\sum_{i=1}^k \epsilon_i$ is small. A detailed comparison between the heuristic and optimal solutions is provided in Fig. \ref{fig11}.

Fig. \ref{fig11} shows that when $\sum_{i=1}^k \epsilon_i$ is above a certain value, our heuristic method actually provides an almost exact optimal solution. The threshold seems roughly proportional to $k \log k$, but more detailed analysis and improvements for the case where $\sum_{i=1}^k \epsilon_i$ is below the threshold will be important issues to be addressed in the future. One possible way to enhance the optimality might be to consider a few more matrix elements for events that more than one attribute values differ.

\ \\
\vspace{-0.5cm}

\begin{figure}[htbt]
  (a) \ \ \ \ \ \ \ \ \ \ \ \ \ \ \ \ \ \ \ \ \ \ \ \ \ \ \ \ \ \ \ \ \ (b)\\
  \centerline{
  \includegraphics[width=4.5cm]{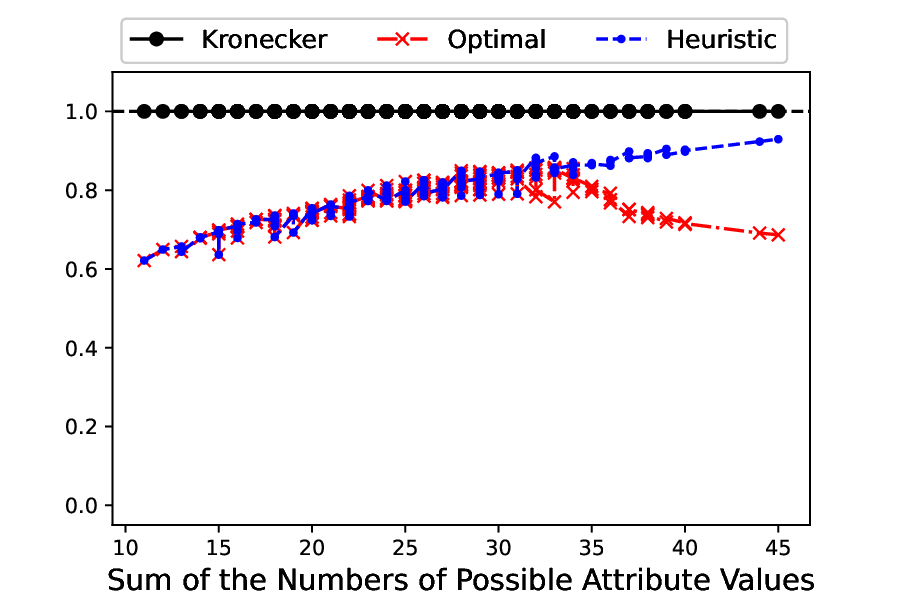}\ 
  \includegraphics[width=4.5cm]{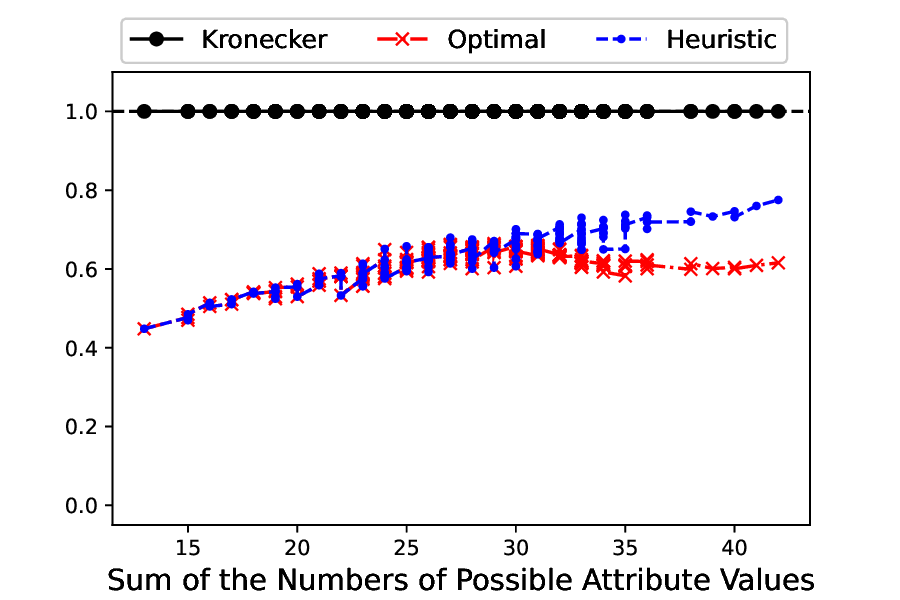}}
  
 (c) \ \ \ \ \ \ \ \ \ \ \ \ \ \ \ \ \ \ \ \ \ \ \ \ \ \ \ \ \ \ \ \ \ (d)\\
  \centerline{
  \includegraphics[width=4.5cm]{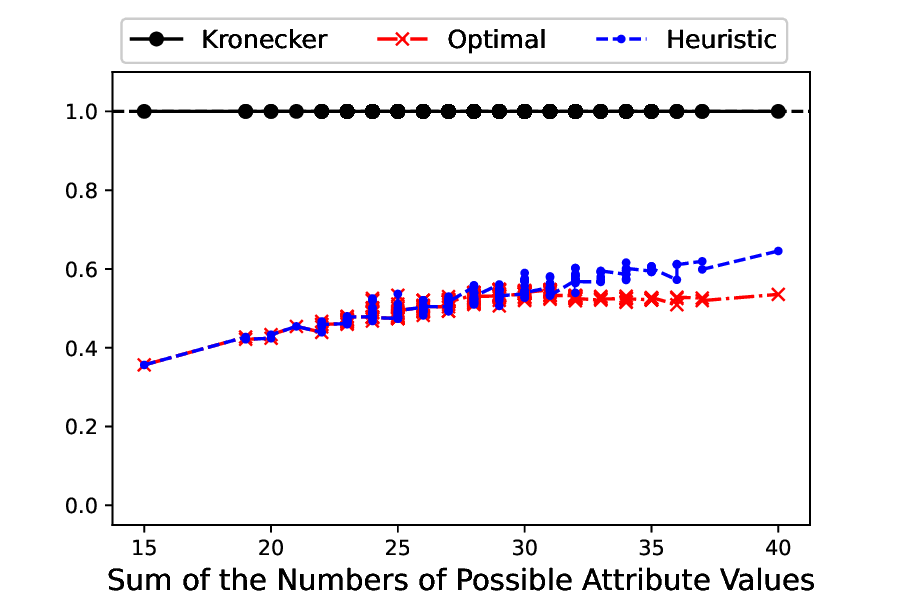}\ \includegraphics[width=4.5cm]{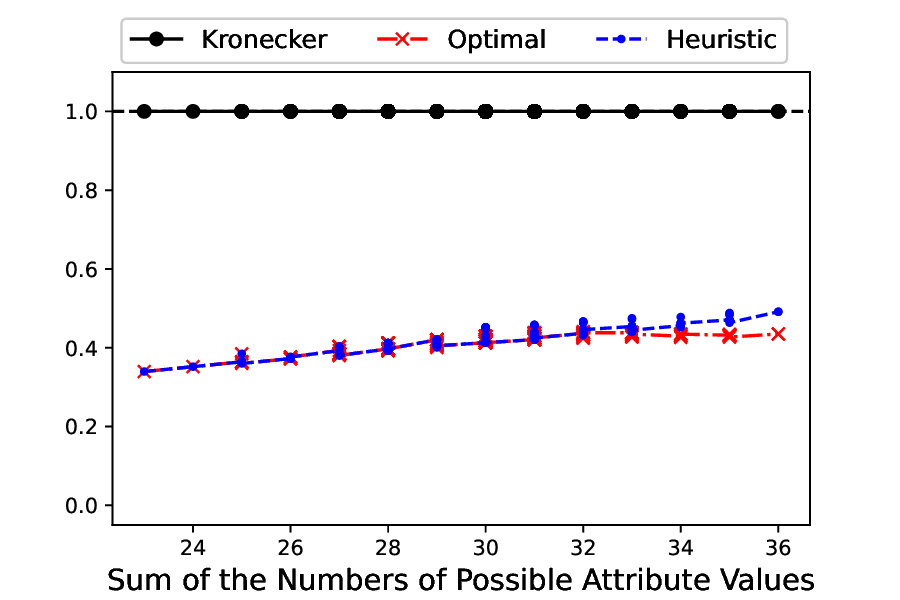}}
  \caption{Achieved privacy levels for the entire dataset when (a) $k = 3$, (b) $k = 5$, (c) $k = 7$, and (d) $k = 10$. The $x$-axis represents $\sum_{i=1}^k a_i$; that is, the sum of the number of possible attribute values for each attribute information. The $y$-axis represents the ratio of the achieved privacy level to $\sum_{i=1}^k \epsilon_i$. We compared the existing Kronecker product-based method (black, solid), the optimal mechanism (red, dash-dot), and our heuristic method (blue, dashed).}
  \label{fig2}
\end{figure}

Next, we show in Fig. \ref{fig2} the results when we fixed $\epsilon_i = 3$ for all $i$. As in the above experiments, for each $k$, the value of each $a_i$ was randomly set and $200$ different sets of $(a_1, \dots, a_k)$ were generated; then, we evaluated the performance of our methods on each set. In the figure, the results are sorted by $\sum_i^k a_i$, which also evaluates the effect of the number of possible values for each attribute information.

Similar to Fig. \ref{fig1}, Fig. \ref{fig2} indicates that our methods can provide significantly stronger privacy guarantees than the existing method. Regarding the impact of $\sum_{i=1}^k a_i$ on the optimal solution, as its value increases, the difference from the Kronecker product-based method becomes smaller at first, but after a certain value, the difference becomes nearly constant or larger. This feature can be discussed using the optimal solution for two-attribute data as in the discussion on Fig. \ref{fig1}.

When $\sum_{i=1}^k a_i$ is small, we can compute the optimal value for cases (I) or (II). As for the privacy level for case (I), the value of (10) tends to increase when the values of $m$ and $n$ increase. However, the increasing trend would become smaller as the sum of privacy level for each attribute information increases. The same is true for case (II).

When $\sum_{i=1}^k a_i$ is large, we can compute the optimal value for cases (III) or (IV). As for the privacy level for case (III), no clear trend in the value of (11) can be observed when the values of $m$ and $n$ vary. The same is true for case (IV).

\begin{figure}[htbt]
  (a) \ \ \ \ \ \ \ \ \ \ \ \ \ \ \ \ \ \ \ \ \ \ \ \ \ \ \ \ \ \ \ \ \ (b)\\
  \centerline{
  \includegraphics[width=4.5cm]{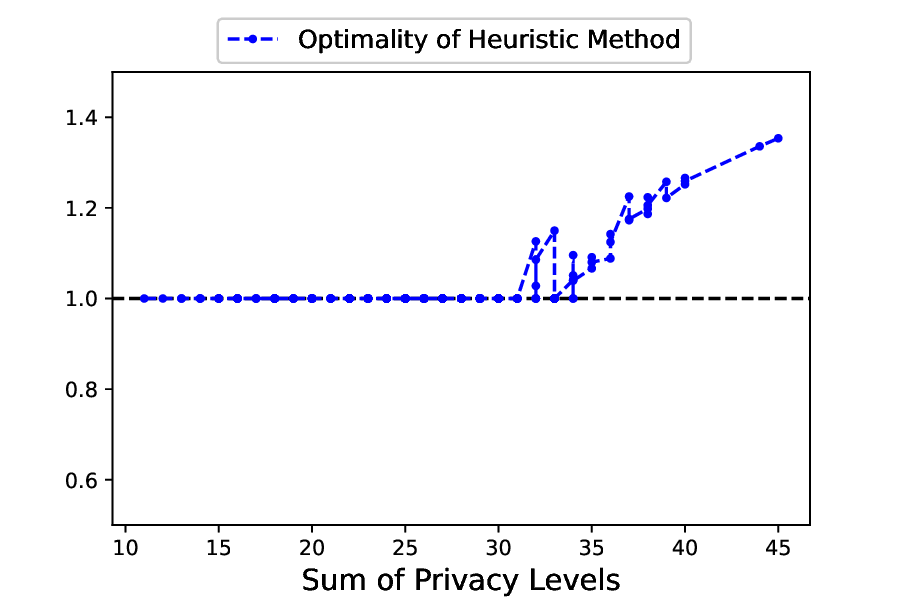}\ 
  \includegraphics[width=4.5cm]{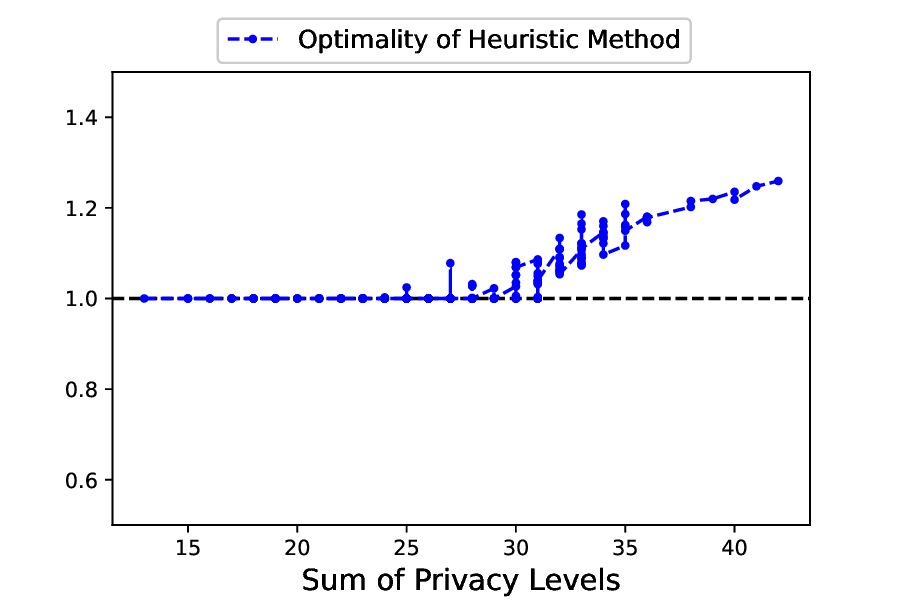}}
  
 (c) \ \ \ \ \ \ \ \ \ \ \ \ \ \ \ \ \ \ \ \ \ \ \ \ \ \ \ \ \ \ \ \ \ (d)\\
  \centerline{
  \includegraphics[width=4.5cm]{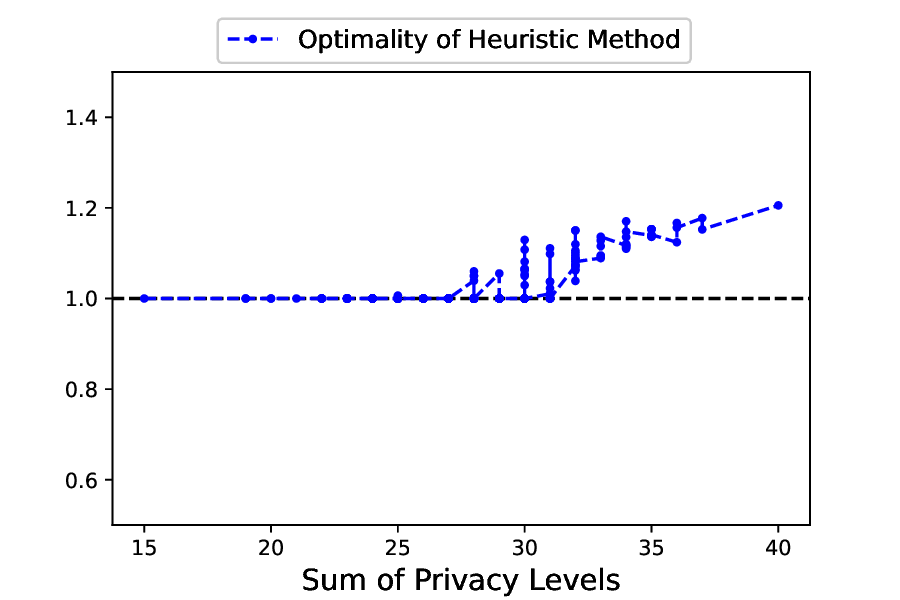}\ \includegraphics[width=4.5cm]{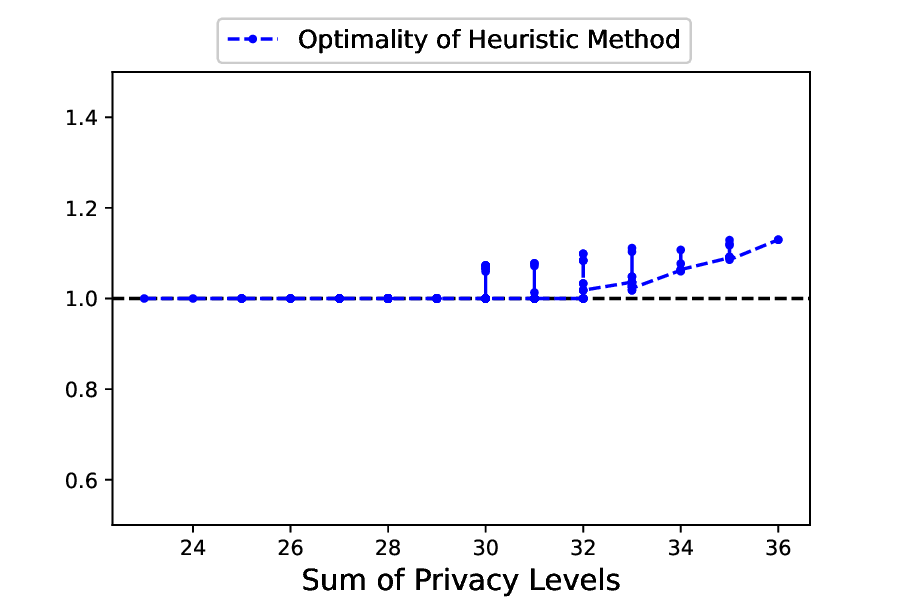}}
  \caption{Optimality of our heuristic method when (a) $k = 3$, (b) $k = 5$, (c) $k = 7$, and (d) $k = 10$. The $x$-axis represents $\sum_{i=1}^k a_i$; that is, the sum of the number of possible attribute values for each attribute information. The $y$-axis represents the ratio of the achieved privacy level to the optimal solution.}
  \label{fig22}
\end{figure}

This feature is expected to be seen for general $k$-attribute data as well and appears in Fig. \ref{fig2}. The finding that the increasing trend becomes smaller as $k$ increases is also predicted from the discussion for case (I). While our heuristic method does not capture this characteristic when $\sum_{i=1}^k a_i$ is large, it can achieve a privacy level close to the optimal solution when $\sum_{i=1}^k a_i$ is small. A detailed comparison is shown in Fig. \ref{fig22}.

Fig. \ref{fig22} shows that when $\sum_{i=1}^k a_i$ is below a certain value, our heuristic method provides an almost exact optimal solution. In this figure, the threshold appears to vary little depending on the value of $k$. This result is consistent with the change in threshold in Fig. \ref{fig11}, both suggesting that when calculating the thresholds, the change in $\epsilon_i$ when $a_i$ is fixed is more sensitive to the change in $k$ than the change in $a_i$ when $\epsilon_i$ is fixed.

\ \\
\vspace{-0.3cm}

In the above, we have shown that when the privacy level for each attribute information is given, the privacy level for the entire dataset can be increased by our methods. This indicates that, conversely, when the overall privacy level is fixed, more privacy budget can be distributed to each attribute information. For example, in case (c) of Fig. \ref{fig1}, if the privacy level for the entire dataset is determined to be $20$, the sum of the budget that can be distributed to each attribute information is also $20$ when using the existing Kronecker product-based method. On the other hand, our methods can distribute a total budget of about $40$. This greatly improves the accuracy of each attribute information after perturbation, which also helps to enhance the accuracy of the data analysis. In the following, we show that our method can provide higher accuracy in data analysis, using genome statistics as an example.

\subsection{Analysis Example}

In this experiment, to verify the utility of our methods on datasets with a larger $k$, we focused on our heuristic method that can be performed in $\mathcal{O}(k^2)$ time and compared its performance with that of the existing method.

In genomic statistical analysis, various statistical tests are often conducted to investigate the relationship between marker loci, such as SNPs, and diseases. The need to protect privacy in the publication of statistics obtained from these tests has been pointed out by several studies \cite{24,25,26}, and the application of differential privacy has recently been well considered \cite{8,27}. Here, we used the most common test, the $\chi^2$-test, as an example to show our method's utility when the privacy level for the entire dataset is fixed. 

Using the information on each SNP $i$, the following $2 \times 2$ contingency table can be constructed:
\vspace{-0.1cm}
\begin{table}[htbt]
    \centering
    \begin{tabular}{cc|cc|c}
        &   & \multicolumn{2}{c|}{Disease Status} & \multirow{2}{*}{Total} \\ \cline{3-4}
        &   & 0 & 1 &  \\ \hline
        \multicolumn{1}{c|}{\multirow{2}{*}{Allele}} & 0 & $A_i$ & $B_i$ & $A_i + B_i$ \\
        \multicolumn{1}{c|}{} & 1 & $C_i$ & $D_i$ & $C_i + D_i$ \\ \hline
        \multicolumn{2}{c|}{Total} & $A_i + C_i$ & $B_i + D_i$ & $2N$
    \end{tabular}
    \vspace{-0.1cm}
\end{table}

\hspace{-0.48cm} where $N$ is the number of individuals. The $\chi^2$-statistic obtained from the table is
\begin{eqnarray}
\frac{2N \cdot (A_i D_i - B_i C_i)^2}{(A_i + B_i)(C_i + D_i)(A_i + C_i)(B_i + D_i)}. \nonumber
\end{eqnarray}
Considering each allele information and disease status as an attribute value, we can employ the randomized response where the number of possible attribute values $(= a_i)$ is $4$ as in the existing study \cite{8}. Then, to share the entire dataset containing such attribute information on $k$ SNPs in total, we can construct a distortion matrix for $k$-attribute data using each privacy level $\epsilon_i$ for the information on SNP $i$ and $a_i = 4$. Here, if the privacy level of the entire dataset is fixed, the value that can be set as $\epsilon_i$ is expected to be larger with our method than with the existing method. Consequently, the accuracy of the $\chi^2$-statistic calculated using the perturbed information is also expected to be higher.

\begin{figure}[htbt]
  \centering
  \includegraphics[width=5.0cm]{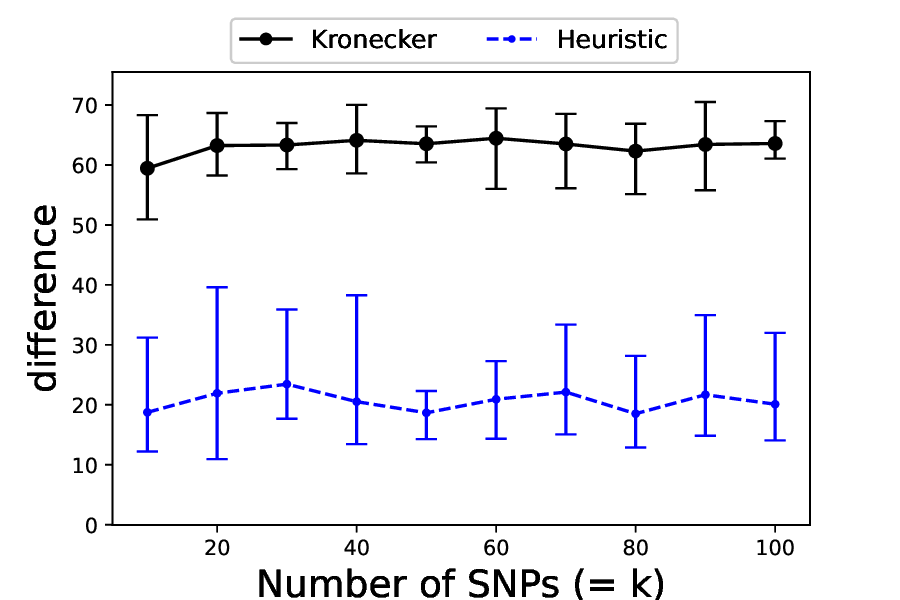}
  \caption{Comparison of the accuracy of $\chi^2$-statistics between the existing Kronecker product-based method (black, solid) and our heuristic method (blue, dashed) when the privacy level for the entire datset is fixed. The $x$-axis represents the number of SNPs; that is, the number of $k$. The $y$-axis represents the average difference between the original and differentially private statistics. The error bar represents the range of all results.}
  \label{fig3}
  \vspace{-0.5cm}
\end{figure}

In the following, we considered the case where the ratio among $\epsilon_i$ $(i = 1,2,\dots,k)$ is fixed, and measured the average difference between the original $\chi^2$-statistic and differentially private statistic after randomized response, while varying the value of $k$. Data for evaluation were generated by setting $N = 1,000$ and the values of $A_i$, $B_i$, $C_i$, and $D_i$ in the contingency table were randomly computed for each SNP $i$ as follows: $A_i = \mathrm{Binomial}(2N, 1/3)$, $B_i = \mathrm{Binomial}(2N - A_i, 1/3)$, $C_i = \mathrm{Binomial}(2N - A_i - B_i, 2/5)$, and $D_i = 2N - A_i - B_i - C_i$. The results over $10$ runs are shown in Fig. \ref{fig3}.

In Fig. \ref{fig3}, the difference when using our heuristic method is almost less than half of that when using the existing method. This result indicates that our method can provide high accuracy in important data analysis. Considering the results of Figs. \ref{fig1} and \ref{fig2}, it is expected that larger values of $\epsilon_i$ or smaller values of $a_i$ will have even increased impact.

\subsection{Run Time}

Furthermore, we measured the run time of our methods for $k$-attribute data. The value of $k$ was varied to $10$ for the optimal method of solving the linear programming problem and up to $1,000$ for the efficient heuristic method. The input values $a_i$ and $\epsilon_i$ were randomly set from $\{2,3,4,5\}$ and $[1,10)$, respectively. The results over $10$ runs are shown in Table \ref{table1}. Note that the values of $a_i$ and $\epsilon_i$ had little effect on the run time of our methods.

\vspace{-0.1cm}
\begin{table}[htbt]
    \centering
    \caption{Run time (sec) of our methods for $k$-attribute data.}
    \vspace{-0.1cm}
    \begin{tabular}{c||c|c|c|c|c|c}
        $k$ & $3$ & $5$ & $10$ & $100$ & $500$ & $1,000$ \\ \hline
        Optimal & $0.0060$ & $0.013$ & $54.9$ & ---- & ---- & ---- \\
        Heuristic & $0.000092$ & $0.00017$ & $0.00051$ & $0.015$ & $0.27$ & $1.14$
    \end{tabular}
    \label{table1}
    \vspace{-0.1cm}
\end{table}

Table \ref{table1} shows that up to about $k = 10$, our optimal method is sufficient to construct the privacy-optimized randomized response in a practical time. In addition, by using our heuristic method, whose time complexity is $\mathcal{O}(k^2)$, we can construct a near-optimal distortion matrix even for datasets with much larger $k$. In fact, the results indicate that for a dataset with $k = 1,000$, our method can be performed in about $1$ second. Moreover, the run time when $k = 10,000$ is less than $2$ minutes, and about $4$ hours when $k = 100,000$. These results are provided on our GitHub page. Coupled with the results in Sections V.A and V.B, we find that our methods can construct an exact or near privacy-optimized randomized response that achieves significantly stronger privacy guarantees for the entire dataset than the existing method for large multi-attribute data within a practical time.

\section{Conclusion}

In this study, we proposed the privacy-optimized randomized response for sharing multi-attribute data, along with a heuristic algorithm for obtaining a near-optimal mechanism in $\mathcal{O}(k^2)$ time. The experimental results confirm that our methods provide significantly stronger privacy guarantees than the existing Kronecker product-based method, leading to higher accuracy in actual data analysis when the privacy level for the entire dataset is fixed. The run time evaluation showed that our heuristic method is feasible in a practical time even for large datasets with $k = 10,000$ or more.

One major limitation of this study is the lack of theoretical guarantees of the heuristic method's optimality. Metaheuristic approaches in addition to inductive methods might be effective, but the development of better heuristic methods is expected to be a more challenging task than it may appear and will be one of the key future issues.

Future research includes developing methods that are specialized for each analysis purpose, as well as exploring methods for datasets that contain numeric data. Furthermore, it would be beneficial to apply our methods to the top-$K$ selection task or to combine them with $k$-anonymization and other important concepts for medical data mining.

\bibliographystyle{plain}
\bibliography{mybibliography}


\end{document}